\documentclass[%
 aip,
 amsmath,amssymb,
 reprint,%
]{revtex4-1}

\usepackage{graphicx}% Include figure files
\usepackage{dcolumn}% Align table columns on decimal point
\usepackage{bm}% bold math
\usepackage[utf8]{inputenc}
\usepackage[T1]{fontenc}
\usepackage{mathptmx}
\usepackage{etoolbox}
\usepackage{amsthm}%
\usepackage{mathrsfs}%
\usepackage{textcomp}%
\usepackage{gensymb}%
\usepackage{xcolor}%
\usepackage{appendix}
\DeclareUnicodeCharacter{2062}{}

\newcommand{\er}{Er$^{3+}$}
\newcommand{\cm}{cm$^{-1}$}
\newcommand{\PTO}{PbTiO$_{3}$}
\makeatletter
\def\@email#1#2{%
 \endgroup
 \patchcmd{\titleblock@produce}
  {\frontmatter@RRAPformat}
  {\frontmatter@RRAPformat{\produce@RRAP{*#1\href{mailto:#2}{#2}}}\frontmatter@RRAPformat}
  {}{}
}%
\makeatother
\begin{document}

\title{Epitaxial Strain Tuning of \er{} in Ferroelectric Thin Films}
\author{Rafaela M. Brinn}
\affiliation{Department of Chemistry, University of California, Berkeley, California 94720, USA}
\author{Peter Meisenheimer}
\affiliation{Department of Materials Science and Engineering, University of California, Berkeley, California 94720, USA}
\author{Medha Dandu}
\affiliation{Molecular Foundry, Lawrence Berkeley National Laboratory, Berkeley, California 94720, USA}
\author{Elyse Barré}
\affiliation{Molecular Foundry, Lawrence Berkeley National Laboratory, Berkeley, California 94720, USA}
\author{Piush Behera}
\affiliation{Department of Materials Science and Engineering, University of California, Berkeley, California 94720, USA}
\author{Archana Raja}
\affiliation{Molecular Foundry, Lawrence Berkeley National Laboratory, Berkeley, California 94720, USA}
\affiliation{Kavli Energy NanoScience Institute, University of California Berkeley, Berkeley, CA 94720}\author{Ramamoorthy Ramesh}
\affiliation{Department of Materials Science and Engineering, University of California, Berkeley, California 94720, USA}
\affiliation{Department of Physics,University of California, Berkeley, California 94720, USA}
\affiliation{Department of Materials Science and Nanoengineering, Rice University, Houston, Texas 77251, USA}
\affiliation{Department of Physics and Astronomy, Rice University, Houston, Texas 77251, USA}
\author{Paul Stevenson}
\affiliation{Department of Physics, Northeastern University, Boston, Massachusetts 02115, USA}
\affiliation{Quantum Materials \& Sensing Institute, Northeastern University, Boston, Massachusetts 02115, USA}
\email{p.stevenson@northeastern.edu }

\date{\today}% It is always \today, today,

\begin{abstract}
\er{} color centers are promising candidates for quantum science and technology due to their long electron and nuclear spin coherence times, as well as their desirable emission wavelength. By selecting host materials with suitable, controllable properties, we introduce new parameters that can be used to tailor the \er{} emission spectrum.  \PTO{} is a well-studied ferroelectric material with known methods of engineering different domain configurations through epitaxial strain. By distorting the structure of \er-doped \PTO{} thin films, we can manipulate the crystal fields around the \er{} dopant. This is resolved through changes in the \er{} resonant fluorescence spectra, tying the optical properties of the defect directly to the domain configurations of the ferroelectic matrix. Additionally, we are able to resolve a second set of peaks for films with in-plane ferroelectric polarization. We hypothesize these results to be due to either the \er{} substituting different sites of the \PTO{} crystal, differences in charges between the \er{} dopant and the original substituent ion, or selection rules. Systematically studying the relationship between the \er{} emission and the epitaxial strain of the ferroelectric matrix lays the pathway for future optical studies of spin manipulation by altering ferroelectric order parameters.
\end{abstract}

\maketitle

\section{Introduction}
Rare-Earth ions (REIs) are attractive photon sources in a broad range of applications, including quantum communication\cite{Stevenson2022Erbium-implantedApplications,Goldner2015RareProcessing,shinErdopedAnataseTiO22022, singhOpticalMicrostructuralStudies2024}and new optical sources\cite{Sun2002RecentApplications,deOliveira2021ADiodes}. The self-contained nature of the 4$f$ electrons\cite{Lucas2015IntroductionMaterials} leads to weak interactions with its environment and makes them attractive systems for storing and manipulating quantum information, demonstrated by impressive coherence properties in both optical and spin degrees of freedom.\cite{Thiel2011Rare-earth-dopedProcessing,Goldner2015RareProcessing,Kunkel2018RecentProcessing,Zhong2019EmergingNanophotonics} In particular, \er{} is a desirable quantum defect due to its electron spin coherence time exceeding 20 ms\cite{LeDantec2021Twenty-threemillisecondCrystal} and nuclear spin coherence times of over a second,\cite{Rancic2018CoherenceMaterial} as well as its wavelength emission in the telecom band ($\sim$1.5 $\mu$m).\cite{Raha2020OpticalQubit} However, in order to fully realize \er-based systems for quantum information sciences, interactions with the host material needs to be carefully considered. \\

Interactions between the \er{} dopant and its host are subtle, yet these interactions are fundamental to our understanding of atomic defects in the solid state. \er{} ions in have been explored for quantum information applications across a wide range of materials and coordination environments, such as yttrium-based crystals\cite{Bottger2006OpticalConcentration,Bottger2009EffectsEr3+:Y2SiO5}, MgO\cite{Baker1976Orbit-latticeStress}, TiO$_2$, \cite{Phenicie2019NarrowTiO2,shinErdopedAnataseTiO22022, singhOpticalMicrostructuralStudies2024} ZnS, PbWO$_4$, MoO$_3$, and ZnO \cite{Stevenson2022Erbium-implantedApplications}. Though the host material is often depicted as a passive environment for the \er{} dopant it can also serve as a resource to provide additional methods of controlling the qubit. From a materials design perspective, this motivates exploration of host materials with additional controllable degrees of freedom (e.g., strain or polarization) to systematically study the dependence of \er{} on its local environment. Important open questions are therefore, in materials with dynamically tunable order parameters, which order parameters can be used to manipulate the defect state and how this can be engineered.

One class of host materials which may offer new functionality for controlling REI quantum defects are ferroelectric materials, where the crystallographic environment can be directly tuned with electric fields and strain. This offers multiple pathways to control the \er{} emission by manipulation of different order parameters \cite{Yadav2016ObservationSuperlattices} or by active modulation of the local coordination environment. Indeed, recent work on {Fe$^{3+}$} doped \PTO{} single crystals has demonstrated that the anisotropy of dopant spins can be controlled by rotation of the ferroelectric polarization in \PTO,\cite{Liu2021CoherentPbTiO3} highlighting the potential for direct coupling between ferroic order and quantum defect states, while piezoelectric materials have been used to coherently drive \er{} defects acoustically.\cite{Ohta2024ObservationIons} In fact, strain has been reported to influence quantum defects by enabling isolating individual REIs in the frequency regime,\cite{Kindem2020ControlCavity} affecting the site symmetry and electric field sensitivity of REIs,\cite{Kindem2020ControlCavity,Shakurov2014RandomCrystals} coupling with quantum emitters to generate entanglement,\cite{Tabesh2021StrainEmitters} and modulating spin relaxation rate in SiV diamond qubits.\cite{Meesala2018StrainDiamond} Therefore, strain manipulation can lead to dynamic coupling and decoupling of ions in ensembles and various control channels.\\

\PTO{} (PTO) is a single composition, ferroelectric perovskite oxide with large spontaneous polarization\cite{Warren1996DefectdipoleFerroelectrics,Ma2014ElectricPbTiO3} and, in thin-films, lattice parameters which can be finely tuned through epitaxy.\cite{Damodaran2017Three-StateFilms} The ability to manipulate the lattice environment with externally-applied fields offers the potential for acoustic control and transduction of quantum states.\cite{Ohta2024ObservationIons,Ashhab2007Two-levelFields,Silveri2017QuantumModulation} Additionally, PTO has a bandgap of $\sim$3.9 eV,\cite{Zelezny2015TheTransition} much greater than the \er{} emission wavelength of ~1.55 eV and facilitating electronic isolation. The particular advantage of functional oxides, like PTO, which have been doped with transition metal ions\cite{Liu2021CoherentPbTiO3} or REIs,\cite{Stevenson2022Erbium-implantedApplications} is that the dopant can have interacting charge, spin, orbital, and/or symmetry orders with the host matrix. Since the distortion of the lattice is directly coupled to the ferroelectric polarization in PTO,\cite{Martin2016Thin-filmApplications} cross coupling between multiple correlated orders can be explored to create solid-state control of quantum emission. This is, however, challenging as deleterious processes such as fast spin relaxation or optical branching ratios are extremely sensitive to the detailed energy level structure. Realizing the full potential of this class of defects in active materials requires a thorough understanding of the interactions between the host lattice and the defect crystal field levels. By utilizing the fine tunability of the lattice parameters of PTO, a systematic study understanding how different crystal fields affect the \er{} optical properties can be performed. \\

In this work, we demonstrate coupling of the \er{} emission to the epitaxial strain of the host matrix up to 80\,K. By systematically changing the epitaxial strain of the host lattice via substrate selection, we modulate both the strain and the crystal field around the defect, allowing exploration of optical properties based on substitutional site occupancy and ferroelectric polarization. Resonant fluorescence spectroscopy is used to study the telecom-wavelength $^4 I _{13/2} \rightarrow ^4 I _{15/2}$ transition for \er, revealing the dependence of peak position, linewidth, and intensity on even subtle changes in epitaxial strain and ferroelectric domain configuration. Understanding the effect of epitaxial strain on \er{} emission provides a blueprint to study the control of \er{} emission with strain using resonant fluorescence spectroscopy.\\

\begin{figure*}
    \includegraphics[width=\textwidth]{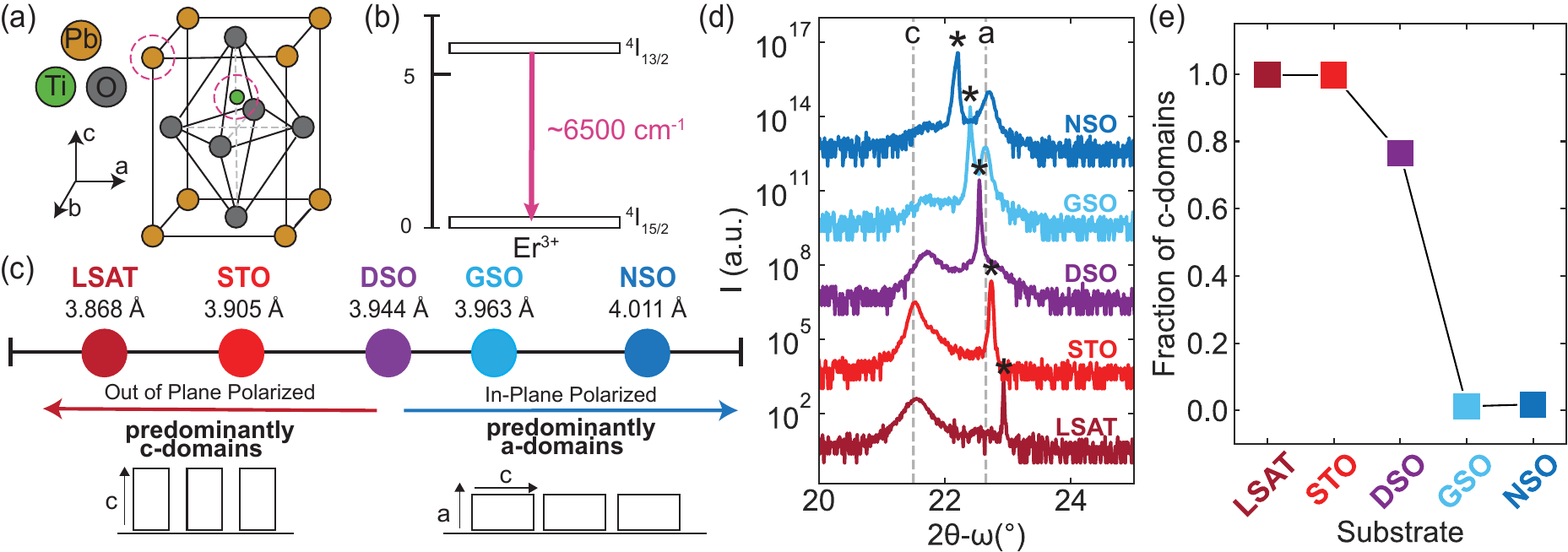}
    \caption{\textbf{\er-doped PTO.} \textbf{a} Diagram of the tetragonal PTO unit cell with the two sites the \er{} dopant can replace (the A- and B-site) outlined in a pink dashed line. \textbf{b} $^4 I _{13/2} \rightarrow ^4 I _{15/2}$ transition of the Er$^{3+}$ ion with an emission of $\sim$6500 \cm{} in the near-IR. \textbf{c} When epitaxially deposited on different substrates, the tetragonal PTO will take on different domain configurations to relax the epitaxial strain of the substrate, ultimately leading to either primarily OOP or primarily IP polarized films. \textbf{d} X-ray diffraction spectra showing the average domain configuration of different films. The peaks at $2\theta - \omega = \sim 21.5$\degree{} and $\sim 22.6$\degree{} correspond to $c$- and $a$-oriented polarizations, respectively. Substrate peak is marked with an *. \textbf{e} This $c:a$ ratio is quantified from the intensity of the peaks in \textbf{d}.}
    \label{domain_config}
\end{figure*}

\section{Results and Discussion}

\subsection{Epitaxial Strain Engineering}

\begin{figure*}
    \includegraphics[width=\textwidth]{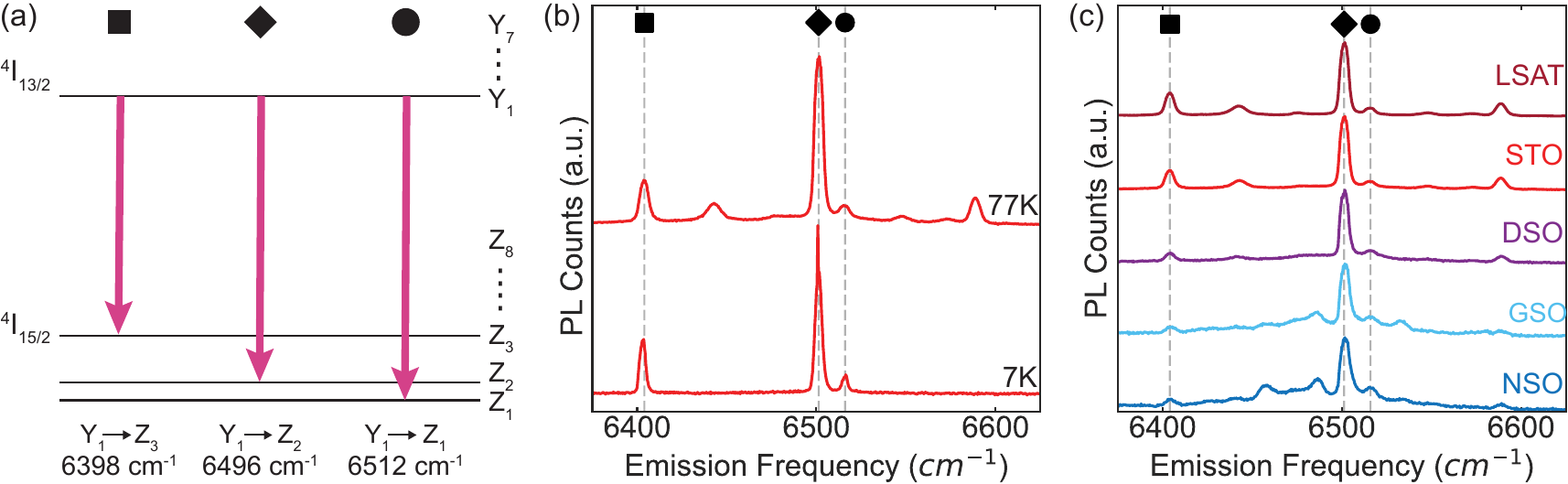}
    \caption{\textbf{Er$^{3+}$ Telecom Transitions.} \textbf{a} Diagram of the energy levels for the $^4 I _{13/2} \rightarrow ^4 I _{15/2}$ transitions. $Z$ and $Y$ refer to the lower and upper manifolds, respectively. \textbf{b} \er{} emission observed for 6500 \cm{} excitation at 77\,K and 6515 \cm{} excitation at 7\,K in \er-doped PTO on STO. \textbf{c} \er{} emission observed for 6500 \cm{} excitation at 77\,K for the 5 films studied here.}
    \label{transitions}
\end{figure*}

To systematically study the effect of strain environment on the \er{} optical properties, \PTO{} is selected as a wide-bandgap, anisotropic host that can be tuned through epitaxial strain. PTO is a tetragonally distorted perovskite (Fig. \ref{domain_config}a) ferroelectric with a significant $c:a$ ($4.11$ \AA$:3.91$ \AA) ratio, where the spontaneous polarization $P$ is along the $c$-axis of the unit cell. This tetragonal distortion of the unit cell translates to a corresponding distortion of the crystal field around the \er{} dopant that can substitute either the A ({Pb$^{2+}$}) or B ({Ti$^{4+}$}) site in the perovskite structure \cite{Dunbar2004ElectronOccupancy} as outlined with a dashed pink line in Figure \ref{domain_config}a. We probe the $^4 I _{13/2} \rightarrow ^4 I _{15/2}$ transition for \er{} at 6500 \cm{} (Fig.  \ref{domain_config}b). Because $c:a$ is large, the orientation of $P$, with respect to the film geometry, can be effectively tuned using thin film heteroepitaxial strain.\cite{Damodaran2017Three-StateFilms}\\

When the lattice constant of the substrate is large ($>3.95$ \AA), creating a tensile epitaxial strain, the PTO film will prefer to form domains where $c$ is in the plane of the film ($a$-oriented) to partially relax the elastic energy. Conversely, when the lattice constant of the substrate is small ($<3.95$ \AA), creating a compressive epitaxial strain, the PTO will form domains where $c$ is preferentially normal to the film plane ($c$-oriented). This is illustrated in Figure \ref{domain_config}c. Because these $a$- and $c$-oriented domains are defined by different lattice constants along the c-axis and a-axis, this difference is apparent in X-ray diffraction (XRD) in Fig. \ref{domain_config}d. As a further complexity, the diffraction peaks are not pinned to the ideal $a$ and $c$ lattice constants, but individual peaks shift betraying further strain within the split domains. Thus varying the substrate lattice parameter varies not only the fraction and orientation of ferroelectric domains, but also the finer strain state within this structure.\\

The fraction of $c$- and $a$-domains can be quantified from XRD using the ratio of the film peaks at 21.5\degree{} ($c$, 001) and 22.6\degree{} ($a$, 100) (Fig. \ref{domain_config}d).\cite{Damodaran2017Three-StateFilms} By integrating the intensity of the $100$- and $001$-peaks, we can quantify the phase fraction of each sample (Fig. \ref{domain_config}e). From this analysis, samples deposited on LSAT and STO substrates, with a compressive epitaxial strain, possess a majority of $c$-domains, while samples on GSO and NSO substrates, under tensile strain, have a majority of $a$-domains. Between these extremes, the lattice constant of DSO is comparable to the mean of the $c$ and $a$ axes of PTO; thus the structure is a mixture of the two configurations. Using piezoresponse force microscopy, we can visualize how the PTO domains change from predominantly $c$-domains in LSAT and STO to predominantly $a$-domains in GSO and NSO, with the DSO sample having some of both domains with a length scales of 10-50 nm (Fig. \ref{PFM}), consistent with the X-ray analysis.\\

\subsection{Emission in Erbium-doped PTO}

 \er{} incorporated within the PTO film can be detected through its luminescence. The \er{} ground state ({$^{4}$I$_{15/2}$}) is composed of eight crystal field levels while the first excited state ({$^{4}$I$_{13/2}$}) is composed of seven. Elevated temperatures result in a highly congested spectrum, expected due to thermalization between states, making it difficult to distinguish and identify because of broad linewidths and thermally excited states. However, at temperatures where the thermal energy, $k_B T$, is small relative to the crystal splittings, a simpler spectrum arising only from the sparser range of thermally-populated states is expected. Analysis here focuses on emission from the {$Y_{1}$} level to the 3 lowest ground states: {$Z_{1}$} (circle), {$Z_{2}$} (diamond), and {$Z_{3}$} (square) (Fig. \ref{transitions}a). Emission is observed that is broadly consistent with previous measurements at 4\,K for \er-doped PTO on SrTiO$_3$ substrate with peaks at 6512 \cm{} ({$Y_{1} \rightarrow Z_{1}$}), 6496 \cm{} ({$Y_{1} \rightarrow Z_{2}$}) and 6398 \cm{} ({$Y_{1} \rightarrow Z_{3}$}),\cite{Stevenson2022Erbium-implantedApplications} though systematic variation of substrate allows identification of new, subtle trends.\\

At 7\,K, only transitions from the {$Y_{1}$} excited state are observed; the long excited state lifetime (3.9 ms)\cite{Stevenson2022Erbium-implantedApplications} ensures the excited state crystal field levels follow a Boltzmann distribution, so only $Y_{1}$ has any appreciable population. At 77\,K, however, additional peaks emerge that correspond to new thermally accessible crystal field levels in the excited state (Fig. \ref{transitions}b). The three transitions from the {$Y_{1}$} level to the the three lowest ground states can be identified for all samples (LSAT, STO, DSO, GSO, NSO) at 77\,K (Fig. \ref{transitions}c). There is, however, a significant variation in peak intensity, the presence of other transitions, and even the frequency of the transitions.\\

\begin{figure*}
    \includegraphics[width=150mm]{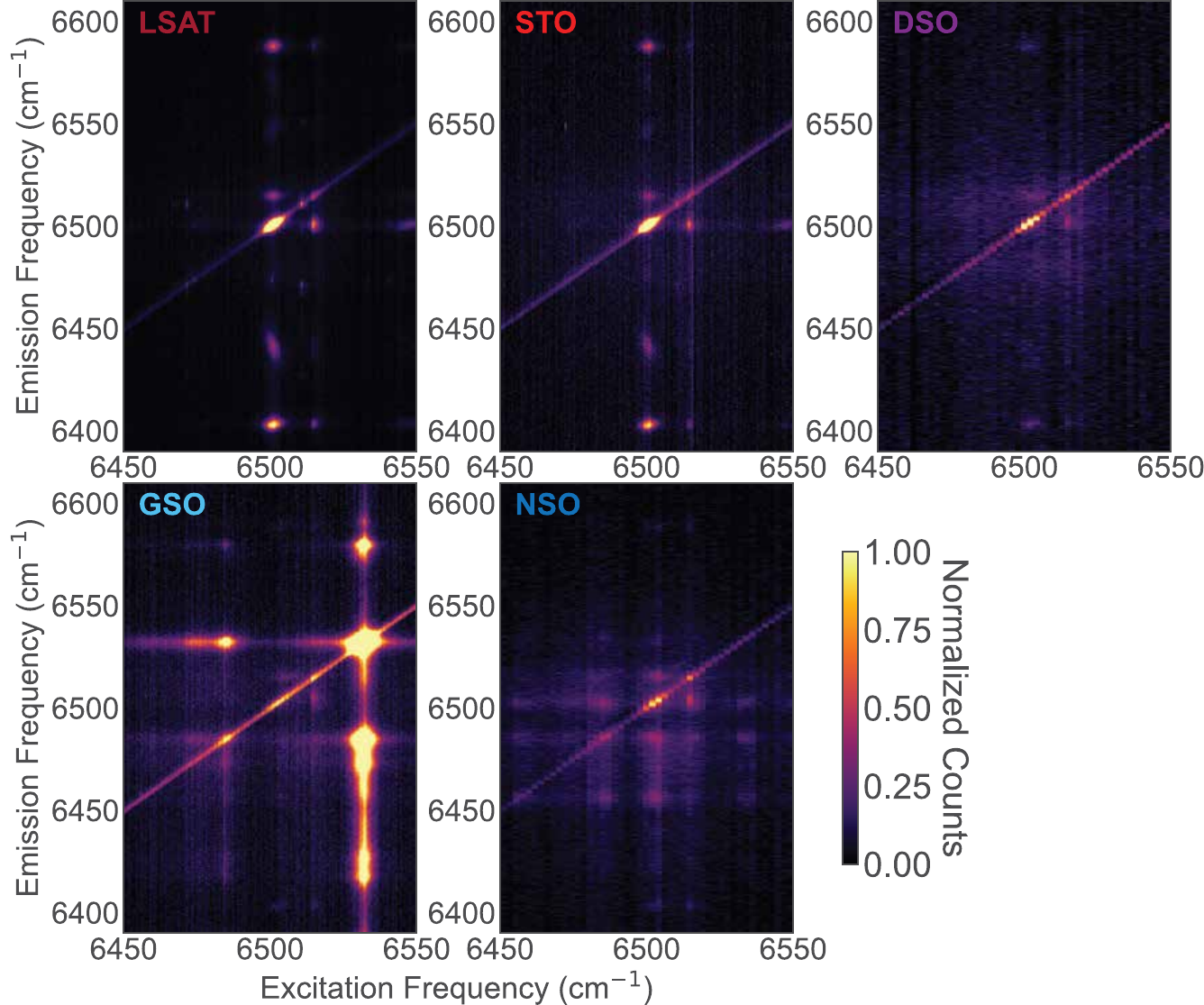}
    \caption{\textbf{Excitation-Emission Spectral Maps.} Spectral maps for all 5 samples. Intensity of spectral maps has been normalized to the 6500 \cm{} emission peak when excited at 6500 \cm. Spectral maps were measured at 77\,K.}
    \label{spectral_maps}
\end{figure*}

To improve our sensitivity to the higher energy peaks and explore internal transitions, excitation-emission spectral maps (Fig. \ref{spectral_maps}) were measured for all five samples of identical thickness. These spectral maps allow visualization of the connection between different excitation and emission peaks. The  diagonal line where emission and excitation frequencies are equivalent corresponds to residual laser scattering. To visualize the change in PL counts from the sample, the intensity is normalized to the 6500 \cm{} emission peak when excited at 6500 \cm{}.\\

From these data, the full complexity of the \er{} photoluminescence can be observed. Emission is observed for all samples at an excitation frequency of 6500 \cm{}. At this excitation frequency, the intensity at other emissions frequencies decreases based on substrate in the following trend: LSAT$>$STO$>$DSO$>$GSO$>$NSO. For the \er-doped PTO sample on GSO, intense emission is detected at excitation frequencies of 6533 \cm{} and 6486 \cm{} which are not observed for the other samples. This emission, however, comes from the substrate itself (discussed later in \textit{Second Site Identification} section), presumably from trace contamination of \er{} in the substrate similar to previous observations of unexpected background \er{} emission.\cite{Wolfowicz2021ParasiticComponents} An additional set of peaks is observed in \er-doped PTO on NSO at an emission frequency of 6457 \cm{} and 6486 \cm{} which are excited by frequencies \textit{not} observed in other samples or in the substrate. From these data, the changes in the \er{} emission as a function of the PTO domain configuration demonstrates intensity, peak position, and linewidth are all impacted by the domain fraction and strain induced by the substrate lattice constant.\\.

\subsection{Strain-Dependent Parameters}
\begin{figure*}
    \includegraphics[width=168mm]{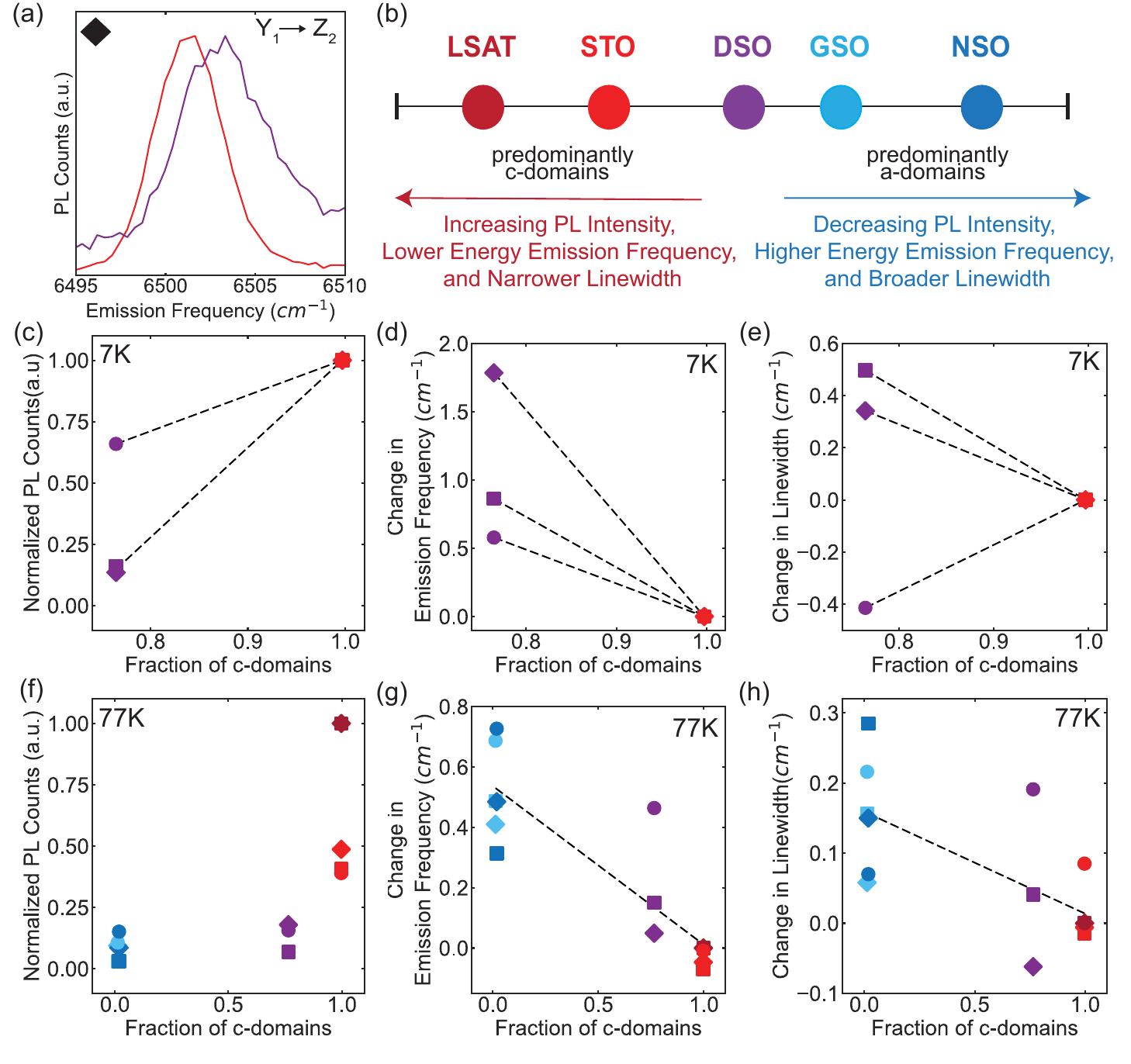}
    \caption {\textbf{Changes in \er{} Emission.} \textbf{a} Emission corresponding to {$Y_{1}$} to  {$Z_{2}$} transitions for \er-doped PTO on STO (red) and DSO (purple) samples at 7\,K. \textbf{b} Diagram of different substrates studied with their observed structural and optical spectral differences. Diagram is also a color reference for the different substrates used in the rest of the figure. \textbf{c} Change in counts, \textbf{d} emission frequency and \textbf{e} peak linewidth for \er-doped PTO sample on STO and DSO at 7\,K. As well as, the \textbf{f} Change in counts, \textbf{g} emission frequency and \textbf{h} peak linewidth for \er-doped PTO sample on the 5 substrates at 77\,K. Normalized PL counts are the PL intensity of each peak normalized to the most intense peak for that transition of all five samples (STO for 7\,K and LSAT for 77\,K). Change in emission frequency and linewidth is compared by subtracting that particular transition with either that of the sample on STO for the 7\,K data set or the sample on LSAT for the 77\,K data set. Measurements done at 7\,K were excited at 6515 \cm{} and measurements done at 77\.K were excited at 6500 \cm. Dashed black trendlines are included to help guide the eye.}
    \label{peak_comparison}
\end{figure*}

First, the difference in intensity, peak linewidth and frequency is visualized by comparing the PL from the \er-doped PTO on STO and DSO samples at 7\,K. The three peaks that are present at 7\,K correspond to the {$Y_{1}$} to {$Z_{1}$} transition (Fig. \ref{PeaksSTODSO7K}a), the {$Y_{1}$} to {$Z_{2}$} transition (Fig. \ref{peak_comparison}a), and the {$Y_{1}$} to {$Z_{3}$} transition (Fig. \ref{PeaksSTODSO7K}b). There is a clear trend in emission across the substrate series (Fig. \ref{peak_comparison}b) consistent across the three peaks. From this we can conclude that the intensity of the peaks increases inversely to the lattice constant, namely NSO$<$GSO$<$DSO$<$STO$<$LSAT, which is the order of $c$-domain fraction. This trend is observed at both 7\,K (Fig. \ref{Raw_PL}a) and 77\,K (Fig. \ref{Raw_PL}b). Changes are also observed in the peak position and linewidth as a function of substrate. Here, the emission peaks shift to higher energies as the fraction of $c$-domains decreases; this trend is clearly observable by inspection at 7\,K, but is also present in the 77\,K data and is accompanied by a broadening of the peaks.\\

By fitting the PL spectra with Gaussian curves with a linear background we can quantify the changes in the emission spectra of the three transitions with fraction of c-domains as a proxy for structural distortion (Fig. \ref{STO_7K}-\ref{NSO_77K}). Differences in PL intensity (Fig. \ref{peak_comparison}c,f), emission frequency (Fig. \ref{peak_comparison}d,g) and peak linewidth (Fig. \ref{peak_comparison}e,h) for the samples at 7\,K (Fig. \ref{peak_comparison}c-e) and 77\,K (Fig. \ref{peak_comparison}f-h) are reported. Fit parameters are reported in Tables \ref{tab:table2} and \ref{tab:table3} for the 7\,K and 77\,K data set, respectively. The PL intensity of the peaks are relative to the most intense peak of that transition (STO for 7\,K (Fig. \ref{peak_comparison}d) and LSAT for 77\,K (Fig. \ref{peak_comparison}g)). Generally, the peak intensities decrease with the decreasing fraction of $c$-domains. At 77\,K, the PL counts decreases on average for the three transitions by 57\% for the STO sample, 86\% for the DSO sample, 92\% for GSO and 91\% for the NSO sample. \\

Additionally, the change in peak position for each transition is compared by subtracting that particular transition with either that of the sample on STO for the 7\,K data set (Fig. \ref{peak_comparison}e) or the sample on LSAT for the 77\,K data set (Fig. \ref{peak_comparison}h). As the fraction of $c$-domains decreases, the emission frequency increases. The change is more drastic for the 7\,K data set where the emission frequency increases by at least 0.58 \cm{} for the {$Y_{1}$} to {$Z_{1}$} transition and at most 1.78 \cm{} for the {$Y_{1}$} to {$Z_{2}$} transitions. At 77\,K, the change in emission frequency for the samples with predominantly $a$-domains vary between 0.31 \cm{} for the {$Y_{1}$} to {$Z_{3}$} transition and 0.73 \cm{} for the {$Y_{1}$} to {$Z_{2}$} transition both for the sample on NSO.\\

Similarly, the change in linewidth for each sample is compared by subtracting each sample's linewidth with either the sample on STO for the 7\,K data set (Fig. \ref{peak_comparison}f) or the sample on LSAT for the 77\,K data set (Fig. \ref{peak_comparison}i). Overall, as the fraction of $c$-domain decreases, the change in linewidth of the samples increases.  The main exceptions is the {$Y_{1}$} to {$Z_{1}$} transition for the \er-doped PTO sample on DSO which changes by -0.41 \cm{} at 7\,K (and is the least intense transition in our dataset). Additionally, the changes in linewidth are smaller than the spectrometer resolution (0.8 \cm), hence these values may be underestimations of the true change in linewidth. Yet again, the change is more drastic at 7\,K than at 77\,K. Overall, the samples with a higher fraction of $c$-domains have brighter peak intensities, lower energy emission frequencies and narrower linewidths. The trends highlighted in Figure \ref{peak_comparison}b demonstrate that even subtle changes in the local environment can tune multiple important emission parameters of the \er{} center. \\

We consider whether these trends could be artifacts from slightly different temperatures due to differences in sample mounting by comparing our data to temperature-dependent measurements (Fig. \ref{Temp_PL}-\ref{Temp_LN2}). In the 10\,K-50\,K range, the PL counts increases for all three peaks up until a threshold temperature of 35\,K is reached in which the PL counts then decrease  and continue to decrease at the liquid N$_2$ temperature regime. This turnover arises from the depopulation of the $Z_1$ ground state once other crystal field levels are thermally excited. The change in PL counts with temperature is much more gradual than the observed changes across samples; for the trends we observe to arise from sample mounting we would require implausibly-large variations of $\approx$25\,K across our datasets. An analysis of the temperature-dependent linewidth yields similar conclusions (see Appendix \ref{A:TDep}).\\

\subsection{Second Site Identification}
\begin{figure*}
    \includegraphics[width=80mm]{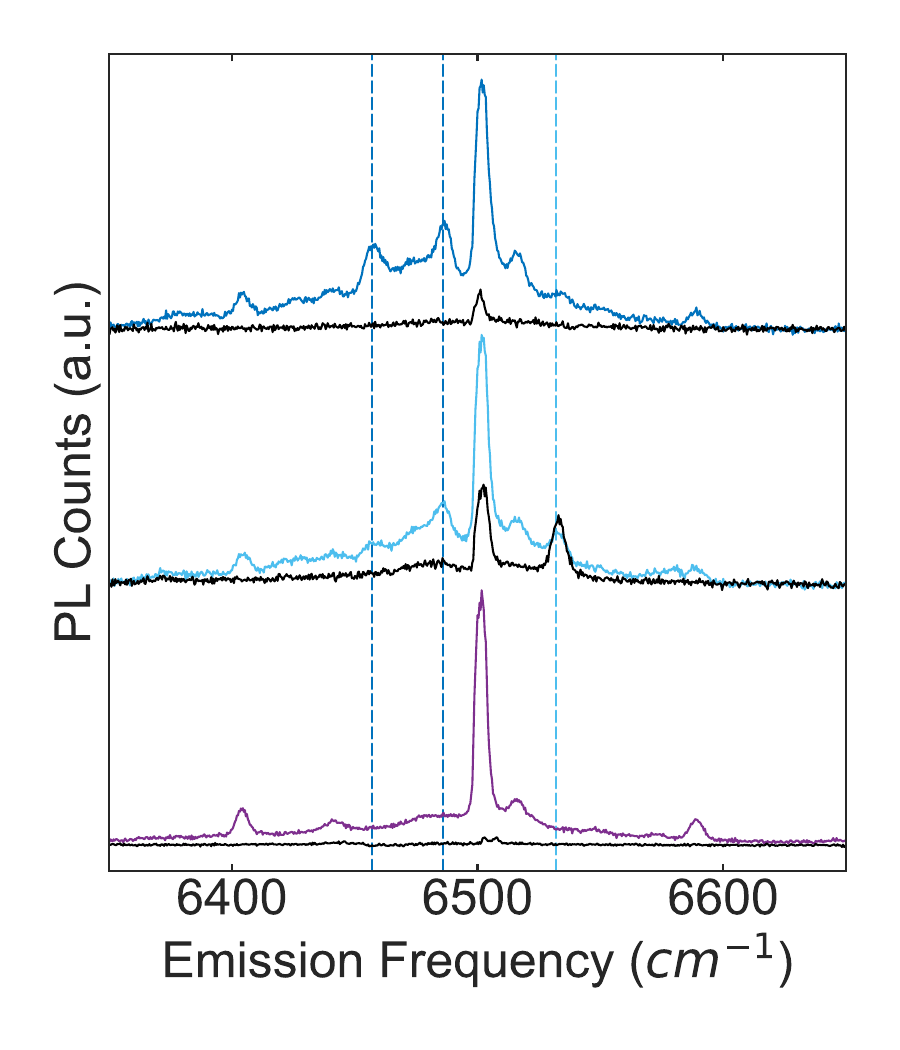}
    \caption{\textbf{Rare Earth Emission from Substrates.} PL for \er{} doped PTO samples deposited on DSO (purple), GSO (light blue) and NSO (dark blue) are compared to that of their respective substrates (black) when excited at 6500 \cm. Substrate spectrum is overlaid on top of the relevant sample spectrum. All spectra are compared relative to their integration time. Dashed lines represent frequencies for different set of peaks present in GSO (light blue dashed) and NSO (dark blue dashed).}
    \label{rare_earth}
\end{figure*}

In addition to the well known spectrum of \er{}, peaks in GSO (Fig. \ref{GSO_newpeaks}) and NSO (Fig. \ref{NSO_newpeaks}) that do not correspond to \er{} emission are observed. In the case of the film on GSO, these emissions arise from the substrate (Fig. \ref{rare_earth} and \ref{REI_subs}a). Given the unique emission range of \er{},\cite{Dieke1963TheEarths} these peaks likely come from \er{} contamination in the substrate itself.\cite{Uda2000TechniqueSeparation,Wolfowicz2021ParasiticComponents} In the case of NSO (Fig. \ref{rare_earth} and \ref{REI_subs}b), however, the emission is not present from the bare substrate, indicating that they arise from the interaction of the doped PTO with the REI-containing substrate.\\

\begin{table*}
\caption{Additional Peaks in \er-doped PTO on GSO and NSO}
\label{tab:table1}
\begin{tabular}{ccc|ccc}
\hline
\multicolumn{3}{c|}{GSO} & \multicolumn{3}{c}{NSO}\\
Frequency  & Intensity  & Linewidth 
& Frequency   & Intensity  & Linewidth \\ 
 (\cm) & (arb.units) & (\cm)
&  (\cm)  & (arb.units) &  (\cm) \\ \hline
6456.43 & 225.48 & 5.06 & 6457.36 & 500.08 & 3.21 \\
6485.71 & 417.52 & 3.01 & 6486.44 & 618.70 & 2.84 \\
6533.38 & 346.06 & 3.07 & 6534.84 & 141.58 & 3.27 \\
\hline
\end{tabular}
\end{table*}

\begin{table*}
\caption{Photoluminescence Excitation Peaks for Different Emission Frequencies for \er-doped PTO on NSO from Figure \ref{spectral_maps}.}
\label{tab:tableExcitation}
\begin{tabular}{c|cccccc}
\hline
\multicolumn{1}{c|}{Emission Frequency} & \multicolumn{5}{c}{Excitation Frequency}\\
\multicolumn{1}{c|}{(\cm)}  & \multicolumn{5}{c}{(\cm)}\\
\hline
6456 & 6457.94 & 6486.18 & 6503.85 & 6515.11 & 6533.21 \\
6485 & 6457.31 & 6485.18 & 6503.17 & 6515.55 & 6532.99 \\
6533 & 6455.68 & 6483.80 & 6503.07 & n/a & 6531.29 \\
\hline
\end{tabular}
\end{table*}

 Alternatively, we hypothesize that there are two different cation sites the \er{} could occupy, each with a different crystal field environment (A site or B site in the ABO$_3$ perovskite structure of \PTO{}) (Fig. \ref{domain_config}a). Since the ionic radius of \er{} (1.03) is relatively similar to Pb$^{2+}$ (1.12) compared to Ti$^{4+}$ (0.56),\cite{Shannon1976RevisedChalcogenides} it is expected for the \er{} dopant to replace the A-site. However, despite the ionic radii mismatch, \er{} has been reported to readily substitute Ti$^{4+}$ in TiO$_2$ in various methods.\cite{Phenicie2019NarrowTiO2,Ji2024Nanocavity-MediatedDeposition,Yang2017StructuralCo-sputtering,Rao2019DefectVOCs,Dibos2022PurcellNanocavities,Sullivan2023Quasi-deterministicControl} Moreover, ESR measurements have shown that REIs can substitute into the A or B site of BaTiO$_{3}$,\cite{Dunbar2004ElectronOccupancy} where the thermodynamic driving force is the ionic radius of the ion relative to the site. This work on BaTiO$_{3}$ is consistent with our observation that the population of these substituent sites changes as we tune the epitaxial strain and lattice parameter of the film with different substrates.\\

A further explanation for the changes in \er{} emission is that the charge compensation required to host \er{} in a lattice of Pb$^{2+}$ and Ti$^{4+}$ changes between substrates due to local elastic energies. Indeed, local \textit{vs} non-local charge compensation, or even different charge compensation configurations, can give rise to distinct sets of peaks.\cite{Tallant1975SelectiveCaF2:Er3+} Vacancies within the PTO crystal (the most likely charge compensation mechanism) have been extensively studied.\cite{Shimada2015MultiferroicSurfaces,Yao2011ChargedPolarization,Serrano2003Oxgen-vacancyStudy} Strain can also affect the concentration of vacancies which, since they provide charge compensation, will impact the  equilibrium concentration of \er{} on either site.\cite{Yang2013TunableStudy} Oxygen vacancies, which are the most common, are highly mobile\cite{Mani2013AtomisticPbTiO3} making them unlikely to be localized around a particular \er{} dopant.\cite{Bredeson2018DimensionalSuperlattices,Zhang2018OxygenSuperlattice}\\

Additionally, changes in sample orientation can result in different selection rules that can alter the \er{} emission spectrum. Different domain configurations affect the local crystal field around the \er{} which is expected to have different optical selection thermodynamics. However, in this case, both set of peaks would be expected on the mixed-phase sample deposited on the DSO substrate which has not been observed. While our spectroscopic results do not allow us to definitely distinguish between these scenarios, we hypothesize that occupancy of the A site \textit{and} B site is the origin of the two sets of peaks.\\

\section{Conclusion}
To conclude, we have shown that \er{} emission can be tuned by epitaxial strain engineering via substrate selection. Epitaxially depositing  \er-doped PTO films at similar conditions on substrates with varying lattice parameters permits fabrication of thin films of the same composition but different domain configurations. This allowed for a systematic comparison of how strain engineering in these films affects the \er{} emission in the spectral range of interest. We observed that films with predominantly $c$-domains have narrower linewidths, emit at lower energies and have a brighter luminescence than films with predominantly $a$-domains. Additionally, samples that have predominantly $a$-domains showed additional peaks that correspond to \er{} transitions. The set of peaks for the \er-doped PTO on GSO sample correspond to \er{} impurities in the GSO substrate itself. However, the set of peaks for the sample on NSO does not come from the NSO substrate and hence must be from the deposited \er-doped PTO. This different set of peaks can either be due to \er{} replacing a different site in the PTO crystal, charge compensation effects, or selection rules. This work lays a foundation to how strain engineering through epitaxial fabrication of samples plays a role in controlling the emission of \er, opening up the pathway to manipulating the properties of REIs via controlling order parameters of the host material.\\

\begin{acknowledgments}
Support for cryogenic optical spectroscopy was provided by the US Department of Energy (DOE) Office of Science, Basic Energy Sciences in Quantum Information Science under award DE-SC0022289. RMB acknowledges fellowship support from the Kavli Philomathia Graduate Student Fellowship. Work at the Molecular Foundry was supported by the Office of Science, Office of Basic Energy Sciences, of the U.S. Department of Energy under Contract No. DE-AC02-05CH11231. RMB would like to thank Edward Barnard, Artiom Skripka, Emory Chang, Daria Blach, Jingxu Xie, and Ari Gashi and the physical infrastructure at the Molecular Foundry for assistance and helpful discussion regarding designing the resonant fluorescence microscope.\\
\end{acknowledgments}

\section*{Data Availability Statement}

The data that support the findings of this study are available from the corresponding author upon reasonable request.

\appendix
\setcounter{figure}{0}
\renewcommand{\thefigure}{A\arabic{figure}}
\setcounter{page}{1}
\renewcommand{\thepage}{A\arabic{page}}
\setcounter{table}{0}
\renewcommand{\thetable}{A\arabic{table}}

\section{Epitaxial PbTiO$_3$ film growth}
100 nm thick thin films of \er-containing PbTiO$_3$ (PTO) were deposited via pulsed laser deposition from an Er containing precursor using $\sim$2 mJ cm$^{-2}$ laser energy at 590 $\degree$C and 100 mTorr process O$_2$. The \er{} concentration is nominally 0.01\% (at.\%). Samples were deposited on substrates with a range of lattice constants in order to generate a wide range of strain environments. The substrates are  (La$_{0.18}$Sr$_{0.82}$)(Al$_{0.59}$Ta$_{0.41}$)O$_3$ (LSAT), SrTiO$_3$ (STO), DyScO$_3$ (DSO), GdScO$_3$ (GSO), and NdScO$_3$ (NSO). 
The strain and domain orientation of the samples were characterized from Cu$K_\alpha$ X-ray diffraction (XRD) spectra about the 001 peak, using the ratio of the $d_c=4.11$ \AA~ and $d_a=3.92$ \AA~ peaks as the $c:a$ domain fraction (Fig. \ref{domain_config}d,e).\\

%\subsection{Resonant Fluorescence Spectroscopy}
%Resonant fluorescence spectroscopy measurements were done at the Molecular Foundry located at the Lawrence Berkeley National Laboratory. Samples were excited using a Newport Velocity Tunable (TLB-6300-LN) laser with a nominal linewidth of $\leq$300 kHz (narrower than any feature discussed here). The excitation passed through a polarizing beam-splitter and the fluorescence with polarization orthogonal to the excitation was detected to suppress scattered laser light. A resonant excitation scheme was designed using Thorlabs Optical Choppers. The emission was dispersed on to an InGaAs AndorCCD camera connected to a Princeton Instruments SpectroPro 300 with a spectral resolution of 0.1 nm (0.8 \cm) (Fig. \ref{laser_table}).

%All measurements were done in a Janis ST-500 cryostat at either 7\,K (liquid He) or 77\,K (liquid N$_2$), with temperature control from a Lakeshore cryogenic temperature controller (Model 325) to enable temperature-dependent measurements.\\

\section{Resonant Fluorescence Spectroscopy}
\begin{figure*}
\includegraphics[width=80mm]{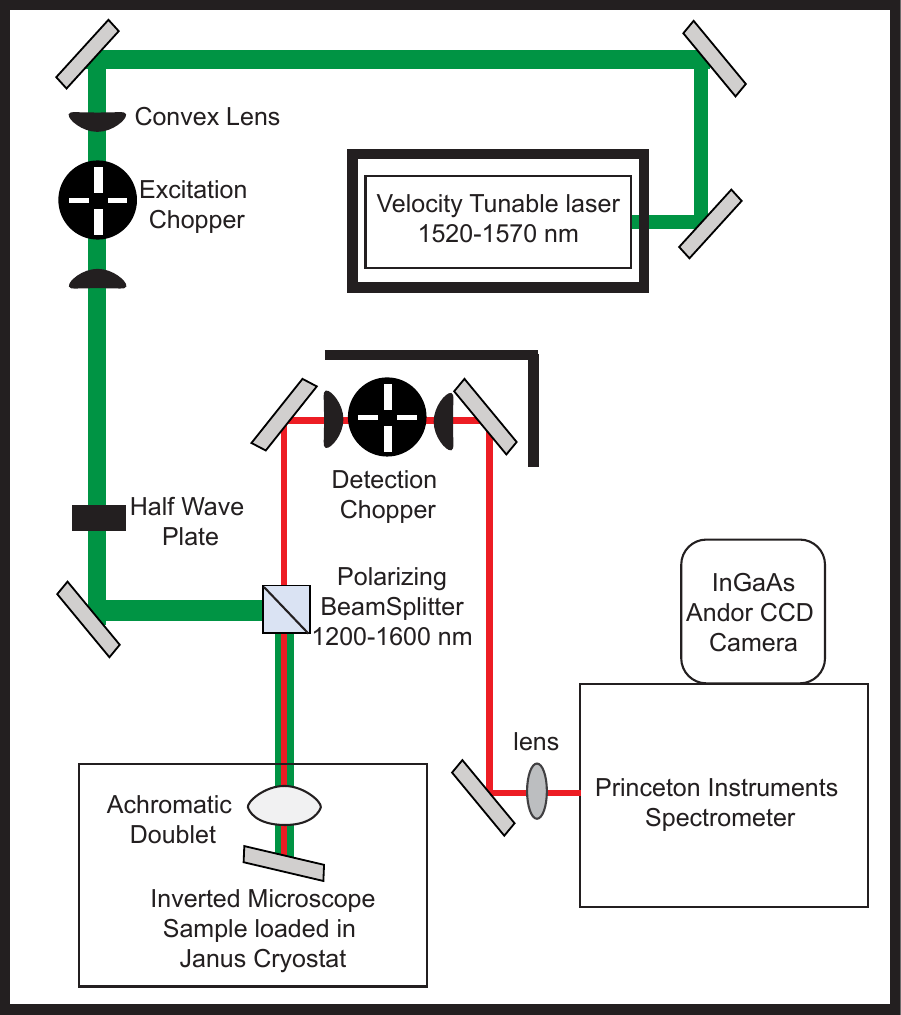}
    \caption{\textbf{Diagram of Resonant Fluorescence Microscope Setup.}}
    \label{laser_table}
\end{figure*}

Resonant fluorescence measurements were done in a home-built microscope set-up (Fig. \ref{laser_table}) constructed at the Molecular Foundry at Lawrence Berkeley National Lab. Samples were excited with a Newport Velocity Tunable (TLB-6300-LN) laser with tunable emission from 1520 nm (6579 \cm)-1570 nm (6369 \cm) and a nominal linewidth of $\leq$300 kHz (narrower than any feature discussed). The excitation passed through a polarizing beam-splitter and the fluorescence with polarization orthogonal to the excitation was detected to suppress scattered laser light. A dual chopper set-up was used to do photoluminescence measurements at resonant excitation-emission frequencies. The choppers were rotating at the same frequency (77 Hz) but with a phase offset to minimize laser scattering. Additional prevention of laser scattering was done by using different chopper blades. The excitation chopper blade only exposed the laser to the sample 10\% of the time while the detection chopper blade only collected the sample emission 50\% of the time. An achromatic doublet lens was used instead of an objective to focus the laser on the sample to image a larger area of the sample. The sample was maintained in either a liquid He or liquid N$_2$ environment under vacuum in a Janis ST-500 cryostat, The emission was dispersed on to an InGaAs AndorCCD camera connected to a Princeton Instruments SpectroPro 300 with a spatial resolution of 0.1 nm (0.8 \cm).\\

\section{PiezoForce Microscopy}
\begin{figure*}
    \includegraphics[width=180mm]{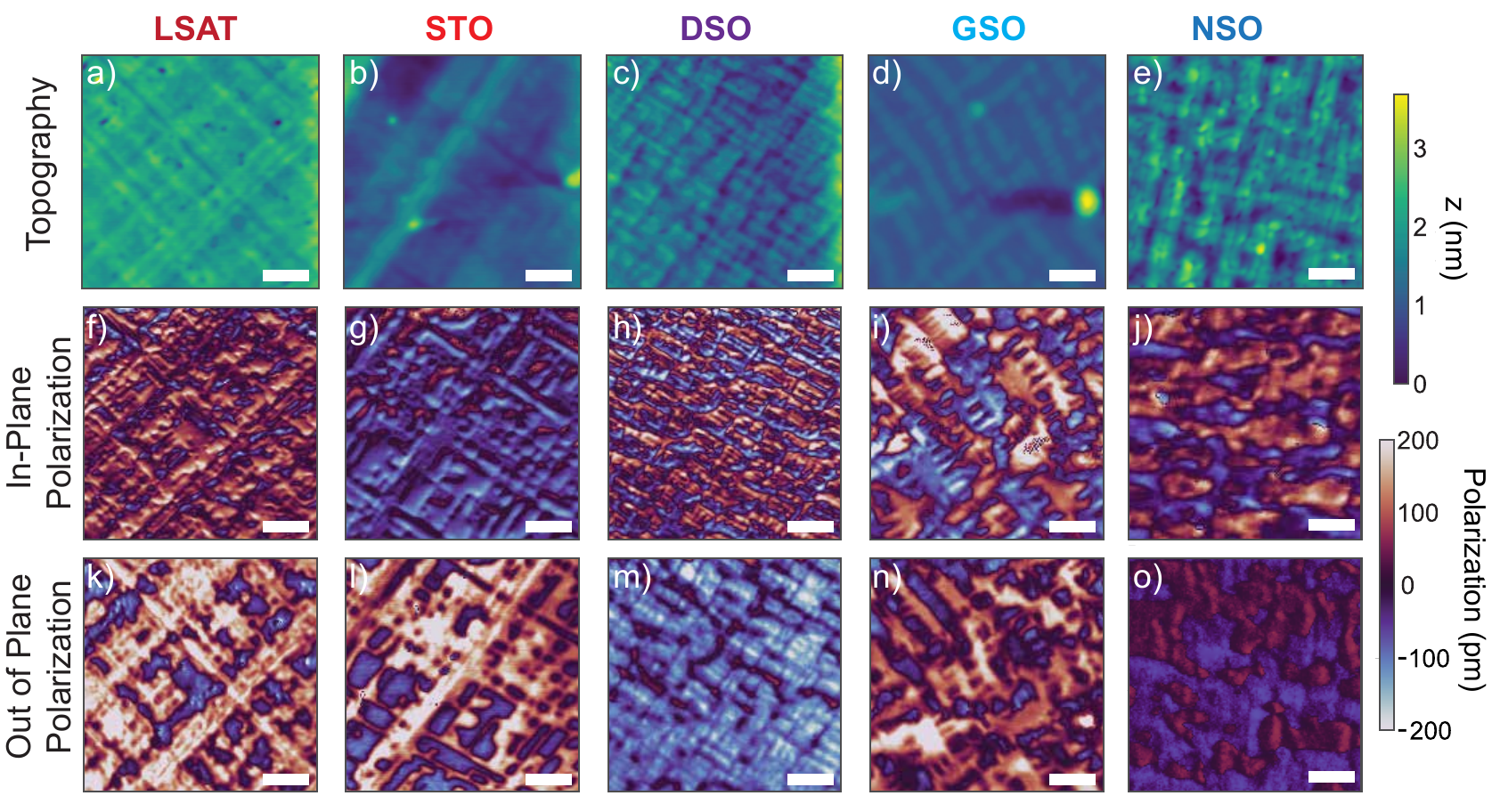}
    \caption{\textbf{Piezoforce Microscopy.}Topography for \er-doped PTO deposited on \textbf{a} LSAT, \textbf{b} STO, \textbf{c} DSO, \textbf{d} GSO, \textbf{e} NSO. All scale bars have a length of 400 nm. Product of amplitude and phase from PFM measurements showing \textbf{f-j} in-plane polarization and \textbf{k-o} out of plane polarization. Corresponding color bars show topographical features (blue to yellow) and polarization domains (blue to red).}
    \label{PFM}
\end{figure*}

Dual AC Resonance Tracking piezoresponse force microscopy (PFM) was done in an atomic force microscope (MFP-3D, Asylum Research). PFM images show the topography (Fig. \ref{PFM}a-e), the in-plane (Fig. \ref{PFM}f-j) and out of plane (Fig. \ref{PFM}k-o) polarizations of the 5 samples. Polarization are represented by the product of the amplitude and phase. The PFM results support the XRD structural characterization of \er-doped PTO. Samples deposited on LSAT and STO exhibit predominantly out-of-plane polarization, while those on GSO and NSO show predominantly in-plane polarization. Samples on DSO exhibit a mixed phase with domain sizes ranging between 10-50 nm.\\

\section{Peak Comparisons}
\begin{figure*}
    \includegraphics[trim={0 16cm 9cm 0},clip, width=150mm]{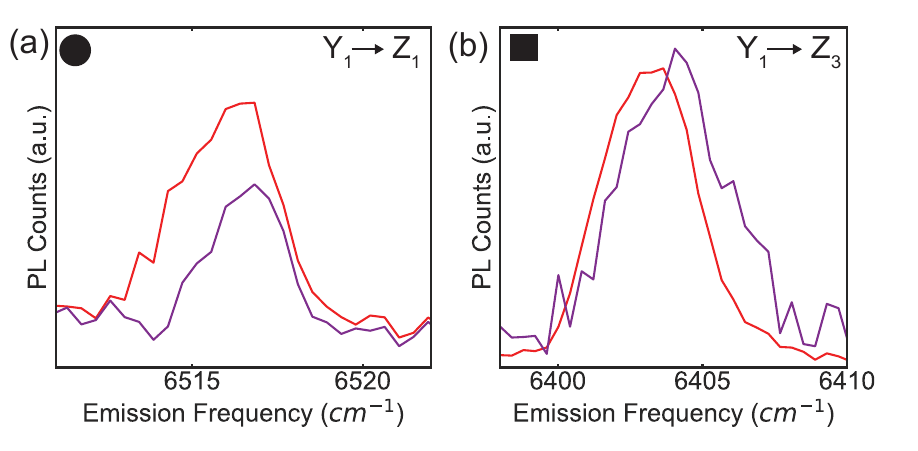}
    \caption{\textbf{Peak Comparison for \er-doped PTO on STO and on DSO at 7K} \textbf{a} Emission corresponding to {$Y_{1}$} to  {$Z_{1}$} and \textbf{b} {$Y_{1}$} to  {$Z_{3}$} transitions for \er-doped PTO on STO (red) and on DSO (purple) samples at 7\,K excited at 6515 \cm. PL was normalized to largest peak intensity.}
    \label{PeaksSTODSO7K}
\end{figure*}

At 7K, the {$Y_{1}$} to  {$Z_{1}$} and {$Y_{1}$} to  {$Z_{3}$} transitions are broader and have a higher energy frequency for \er-doped PTO deposited on DSO than on STO.\\

\begin{figure*}
    \includegraphics[width=150mm]{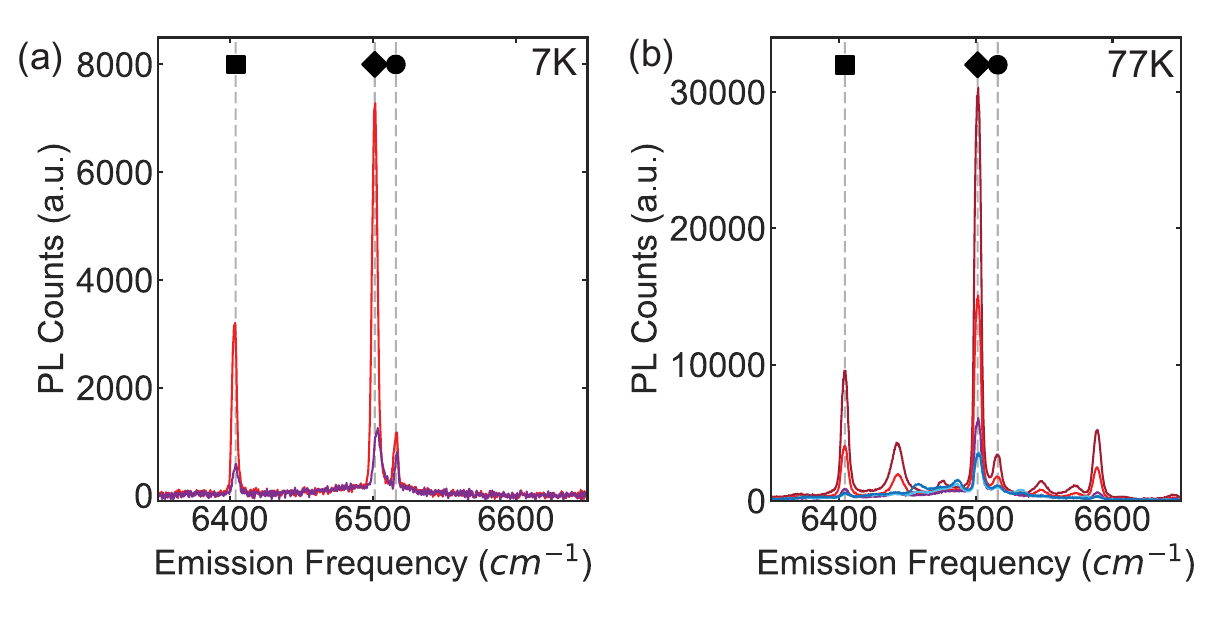}
    \caption{\textbf{Relative Intensity of \er-doped PTO.} \textbf{a} Relative PL spectra of \er-doped PTO on STO (red) and on DSO (purple) at 7\,K excited at 6515 \cm{} and \textbf{b} Relative PL spectra of \er-doped PTO on LSAT (dark red), STO (red), DSO (purple), GSO (light blue), NSO (dark blue) at 77\,K excited at 6500 \cm.}
    \label{Raw_PL}
\end{figure*}

Relative intensities of the photoluminescence (PL) of \er-doped PTO deposited in different substrates at 7\,K (Fig. \ref{Raw_PL}a) and 77\,K (Fig. \ref{Raw_PL}b). At both temperatures, samples with a higher fraction of $c$-domains have a greater number of PL counts.\\

\subsection{Peak Fits for \er-doped PTO Samples at 7\,K}
\begin{figure*}
    \includegraphics[width=142mm]{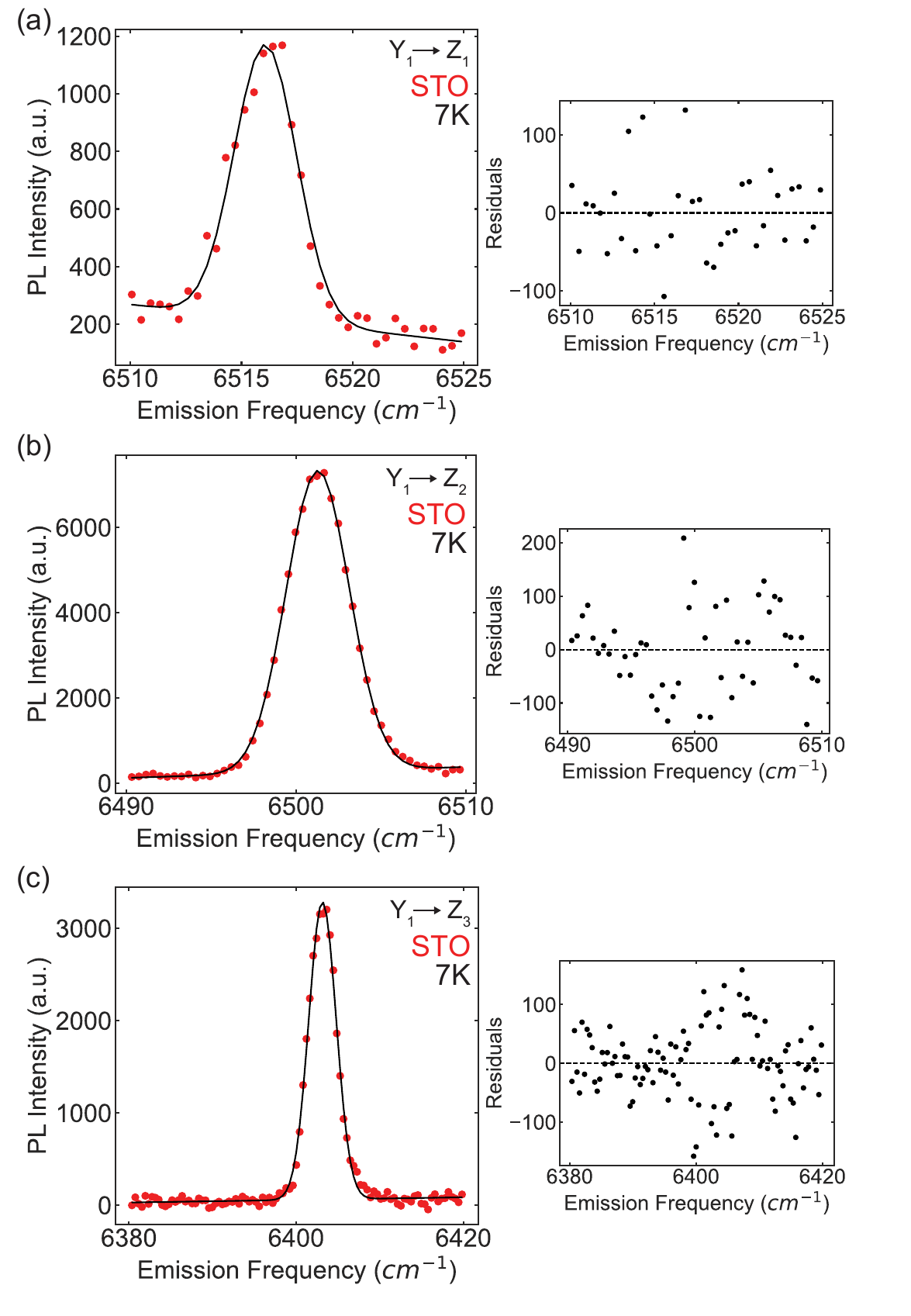}
    \caption{\textbf{Fits for \er-doped PTO on STO at 7K.} Gaussian fits for the \textbf{a} {$Y_{1} \rightarrow Z_{1}$}, \textbf{b} {$Y_{1} \rightarrow Z_{2}$}, and \textbf{c} {$Y_{1} \rightarrow Z_{3}$} transitions. Fits (black solid line) shown on top of data (red circle). Corresponding residual from fits shown to the right of each plot. Sample was excited at 6515 \cm.}
    \label{STO_7K}
\end{figure*}

\begin{figure*}
    \includegraphics[width=142mm]{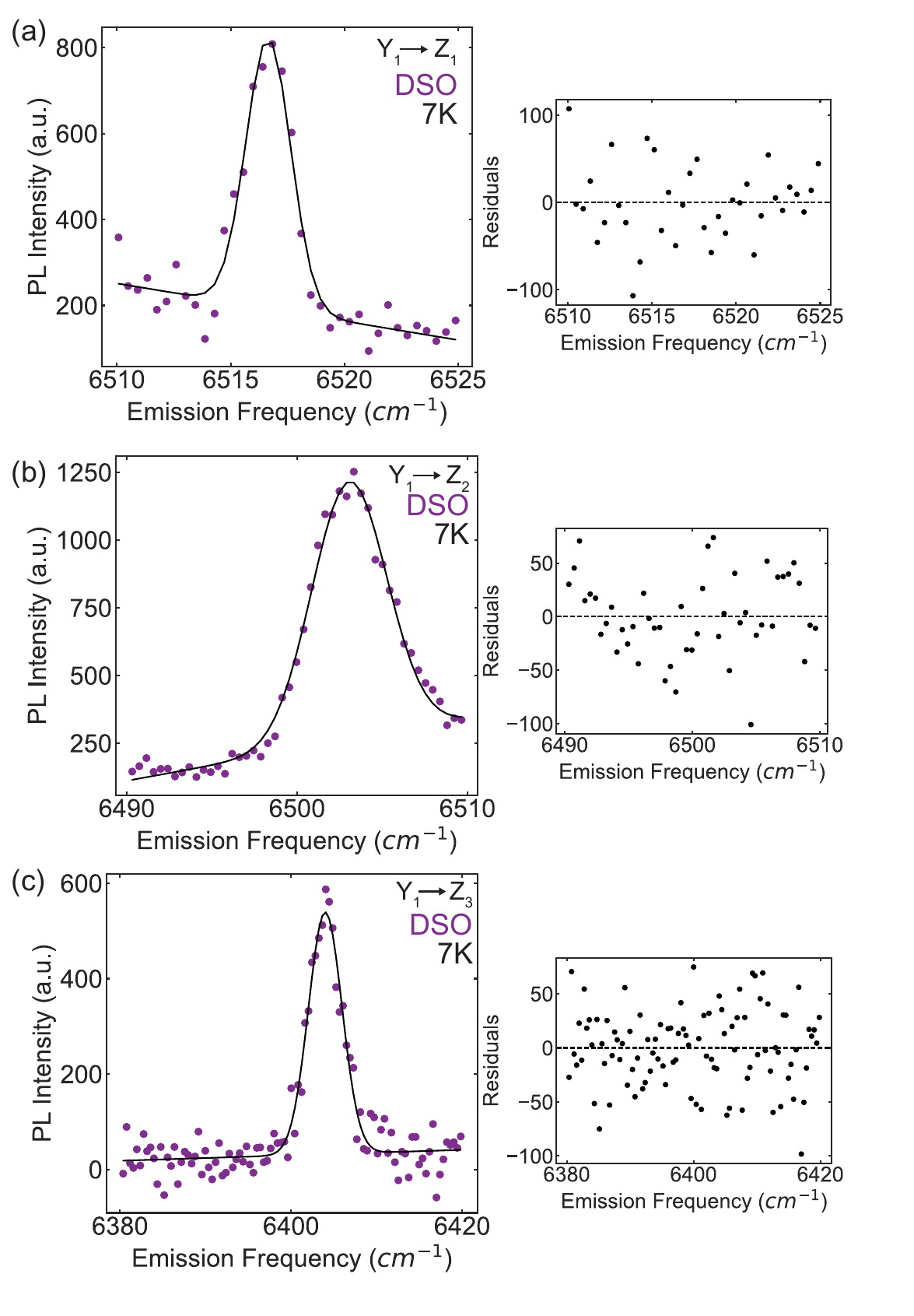}
    \caption{\textbf{Fits for \er-doped PTO on DSO at 7\,K.} Gaussian fits for the \textbf{a} {$Y_{1} \rightarrow Z_{1}$}, \textbf{b} {$Y_{1} \rightarrow Z_{2}$}, and \textbf{c} {$Y_{1} \rightarrow Z_{3}$} transitions. Fits (black solid line) shown on top of data (purple circle). Corresponding residual from fits shown to the right of each plot. Sample was excited at 6515 \cm}
    \label{DSO_7K}
\end{figure*}

Peak fits for the \er-doped PTO samples on STO (Fig. \ref{STO_7K}) and DSO (Fig. \ref{DSO_7K}) at 7\,K. Peaks were fit to a Gaussian curve with a linear background. The three peaks fitted correspond to the {$Y_{1} \rightarrow Z_{1}$} (Fig. \ref{STO_7K}a and \ref{DSO_7K}a), {$Y_{1} \rightarrow Z_{2}$} (Fig. \ref{STO_7K}b and \ref{DSO_7K}b), and  {$Y_{1} \rightarrow Z_{3}$} (Fig. \ref{STO_7K}c and \ref{DSO_7K}c) transitions. Residual of fits provided next to each peak fit to show the goodness of fits. Fit results are reported in Table \ref{tab:table2}.\\

\begin{table*}
\caption{Gaussian fit results of frequency (Freq.), intensity, and linewidth from Figures \ref{STO_7K} and \ref{DSO_7K}. Freq. from Literature (Lit.) from \cite{Stevenson2022Erbium-implantedApplications}.}
\label{tab:table2}
\begin{tabular}{cc|ccc|ccc}
\hline
 & & \multicolumn{3}{c}{STO}&\multicolumn{3}{c}{DSO}\\
Transition & Lit. value & Freq. & Intensity & Linewidth & Freq. & Intensity & Linewidth\\
&(\cm) & (\cm) & (arb. units) & (\cm) & (\cm) & (arb. units) & (\cm)\\ \hline
{$Y_{1} \rightarrow Z_{1}$} & 6511.7 & 6516.08 & 954.96 & 1.4 & 6516.66 & 629.35 & 0.99\\
{$Y_{1} \rightarrow Z_{2}$}& 6497.49 & 6501.27 & 7066.52 & 1.86 & 6503.06 & 954.91 & 2.2\\
{$Y_{1} \rightarrow Z_{3}$}& 6398.23 & 6403.15 & 3224.14 & 1.64 & 6403.99 & 507.47 & 1.96\\
\hline
\end{tabular}
\end{table*}

\subsection{Peak Fits for \er-doped PTO Samples at 77\,K.}
\begin{figure*}
    \includegraphics[width=142mm]{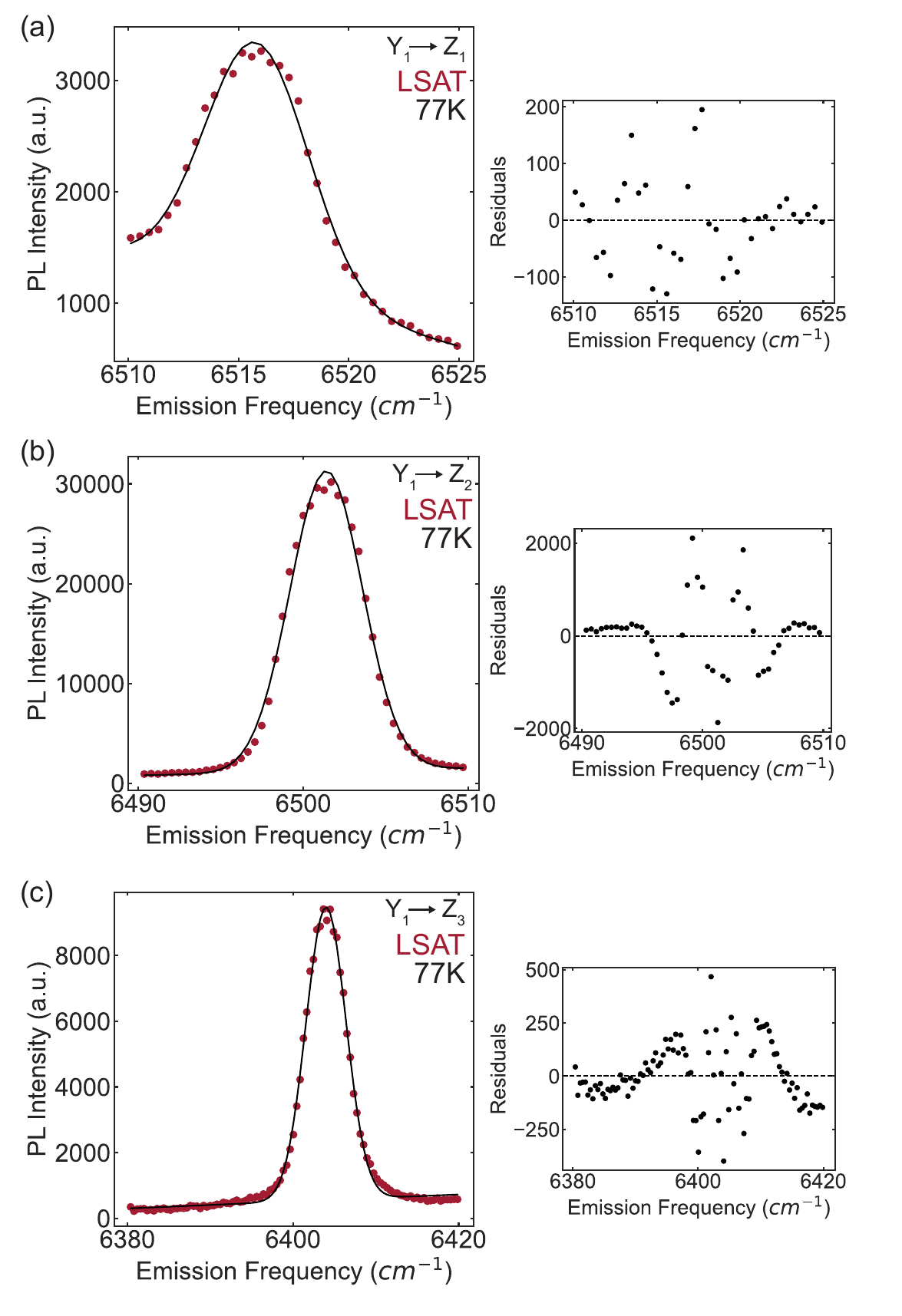}
    \caption{\textbf{Fits for \er-doped PTO on LSAT at 77K.} Gaussian fits for the \textbf{a} {$Y_{1} \rightarrow Z_{1}$}, \textbf{b} {$Y_{1} \rightarrow Z_{2}$}, and \textbf{c} {$Y_{1} \rightarrow Z_{3}$} transitions. Fits (black solid line) shown on top of data (dark red circle). Corresponding residual from fits shown to the right of each plot. Sample was excited at 6500 \cm.}
    \label{LSAT_77K}
\end{figure*}

\begin{figure*}
    \includegraphics[width=142mm]{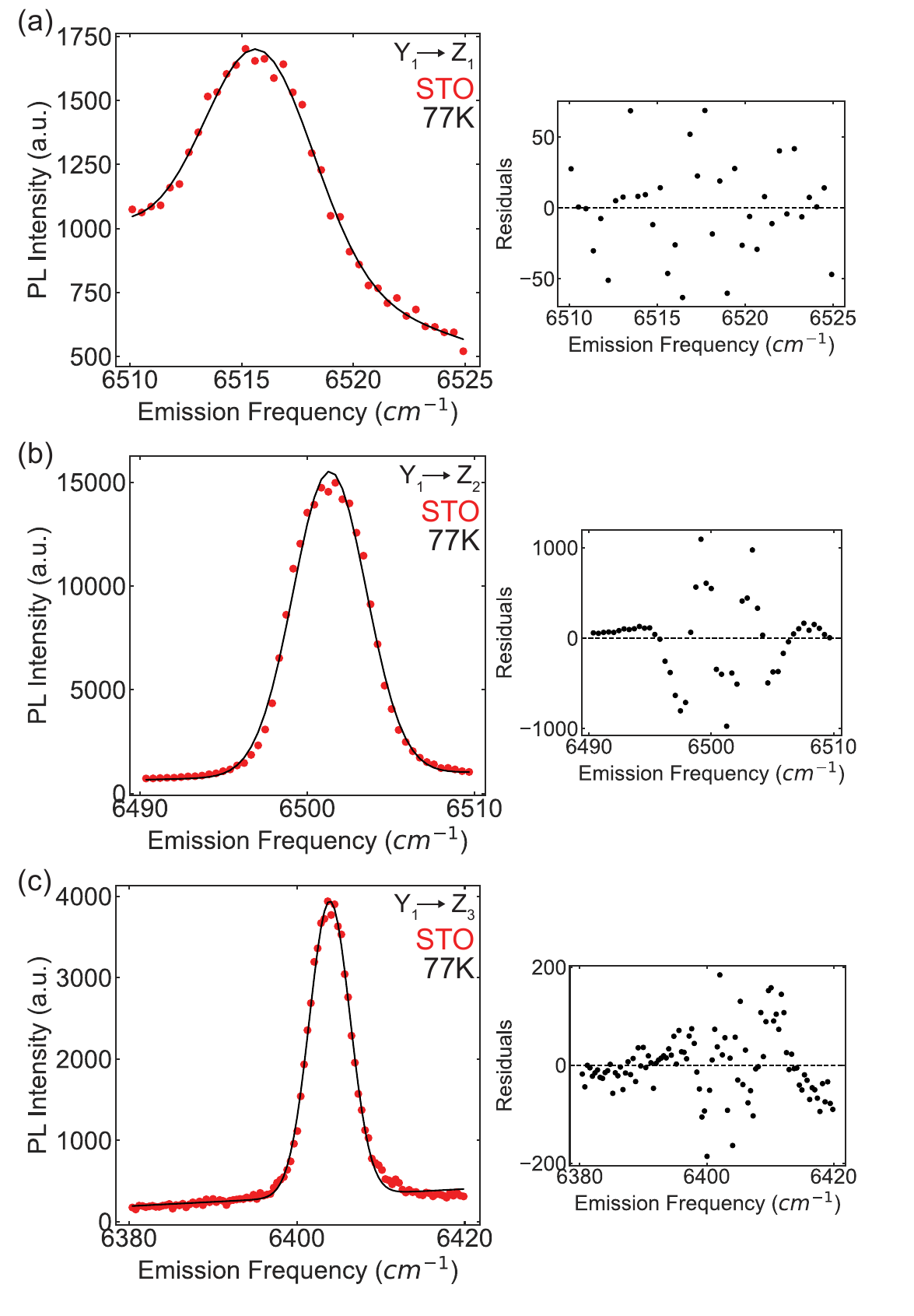}
    \caption{\textbf{Fits for \er-doped PTO on STO at 77\,K.} Gaussian fits for the \textbf{a} {$Y_{1} \rightarrow Z_{1}$}, \textbf{b} {$Y_{1} \rightarrow Z_{2}$}, and \textbf{c} {$Y_{1} \rightarrow Z_{3}$} transitions. Fits (black solid line) shown on top of data (red circle). Corresponding residual from fits shown to the right of each plot. Sample was excited at 6500 \cm.}
    \label{STO_77K}
\end{figure*}

\begin{figure*}
    \includegraphics[width=70mm]{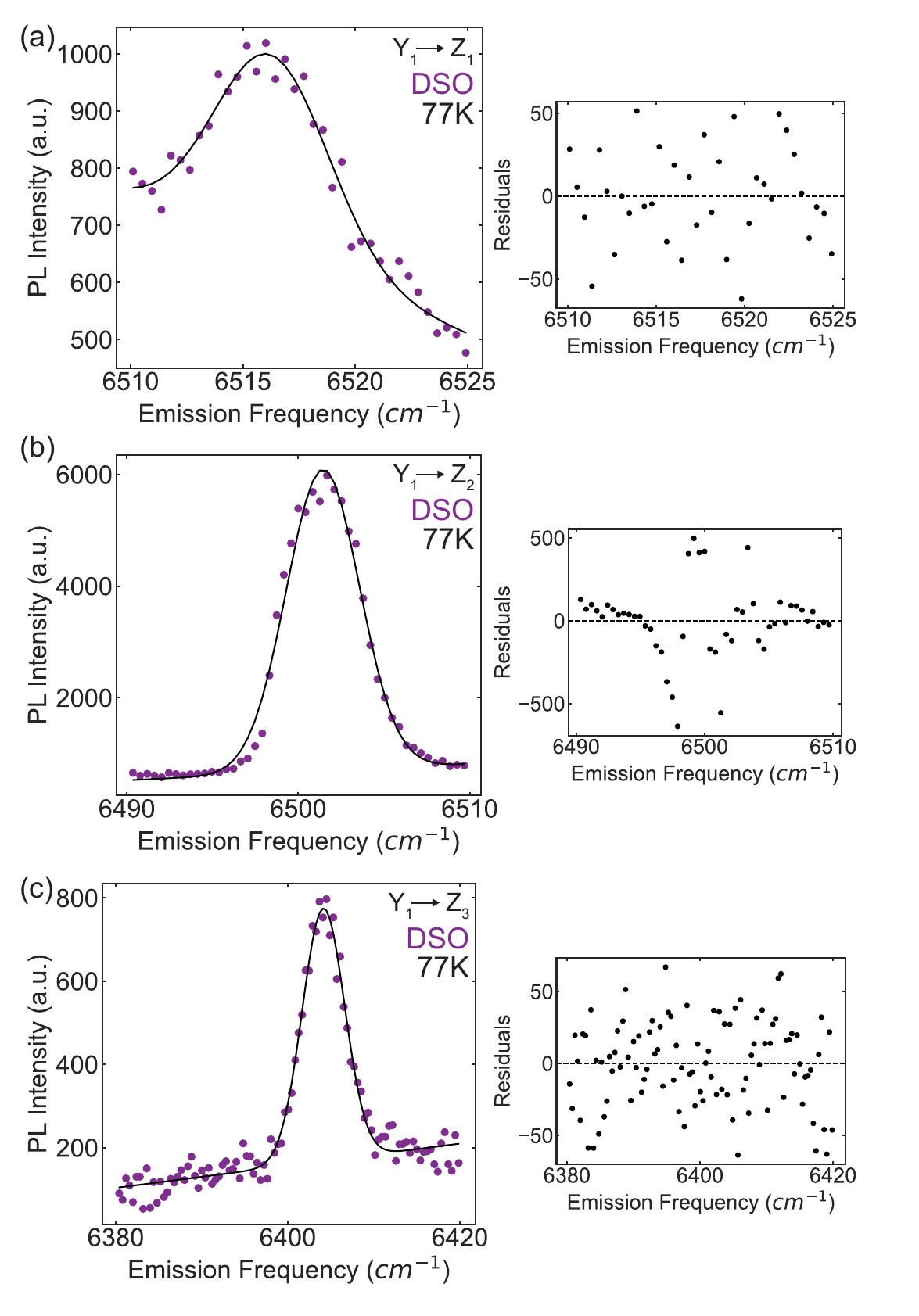}
    \caption{\textbf{Fits for \er-doped PTO on DSO at 77\,K.} Gaussian fits for the \textbf{a} {$Y_{1} \rightarrow Z_{1}$}, \textbf{b} {$Y_{1} \rightarrow Z_{2}$}, and \textbf{c} {$Y_{1} \rightarrow Z_{3}$} transitions. Fits (black solid line) shown on top of data (purple circle). Corresponding residual from fits shown to the right of each plot. Sample was excited at 6500 \cm.}
    \label{DSO_77K}
\end{figure*}

\begin{figure*}
    \includegraphics[width=70mm]{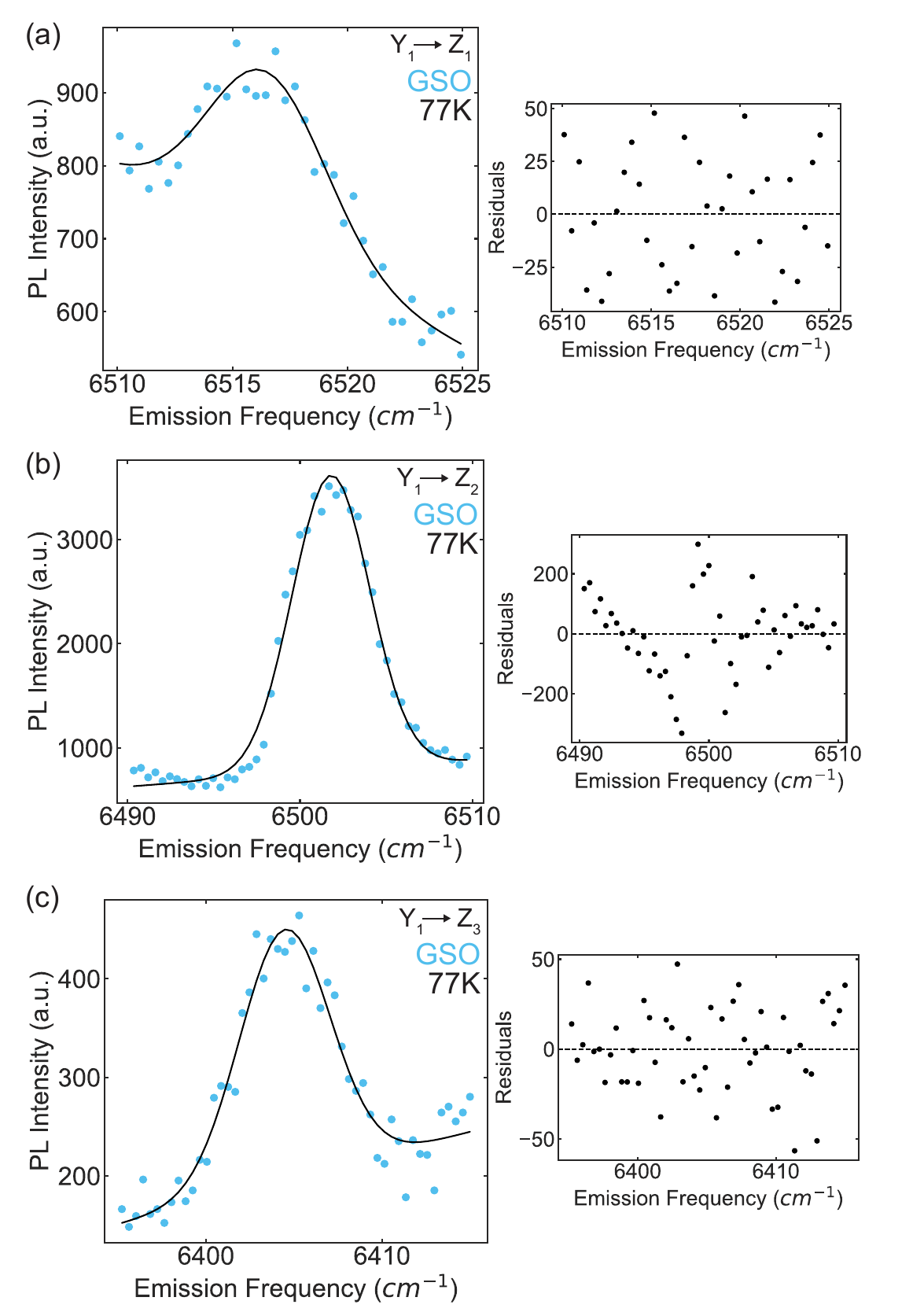}
    \caption{\textbf{Fits for \er-doped PTO on GSO at 77\,K.} Gaussian fits for the \textbf{a} {$Y_{1} \rightarrow Z_{1}$}, \textbf{b} {$Y_{1} \rightarrow Z_{2}$}, and \textbf{c} {$Y_{1} \rightarrow Z_{3}$} transitions. Fits (black solid line) shown on top of data (light blue circle). Corresponding residual from fits shown to the right of each plot. Sample was excited at 6500 \cm.}
    \label{GSO_77K}
\end{figure*}

\begin{figure*}
    \includegraphics[width=70mm]{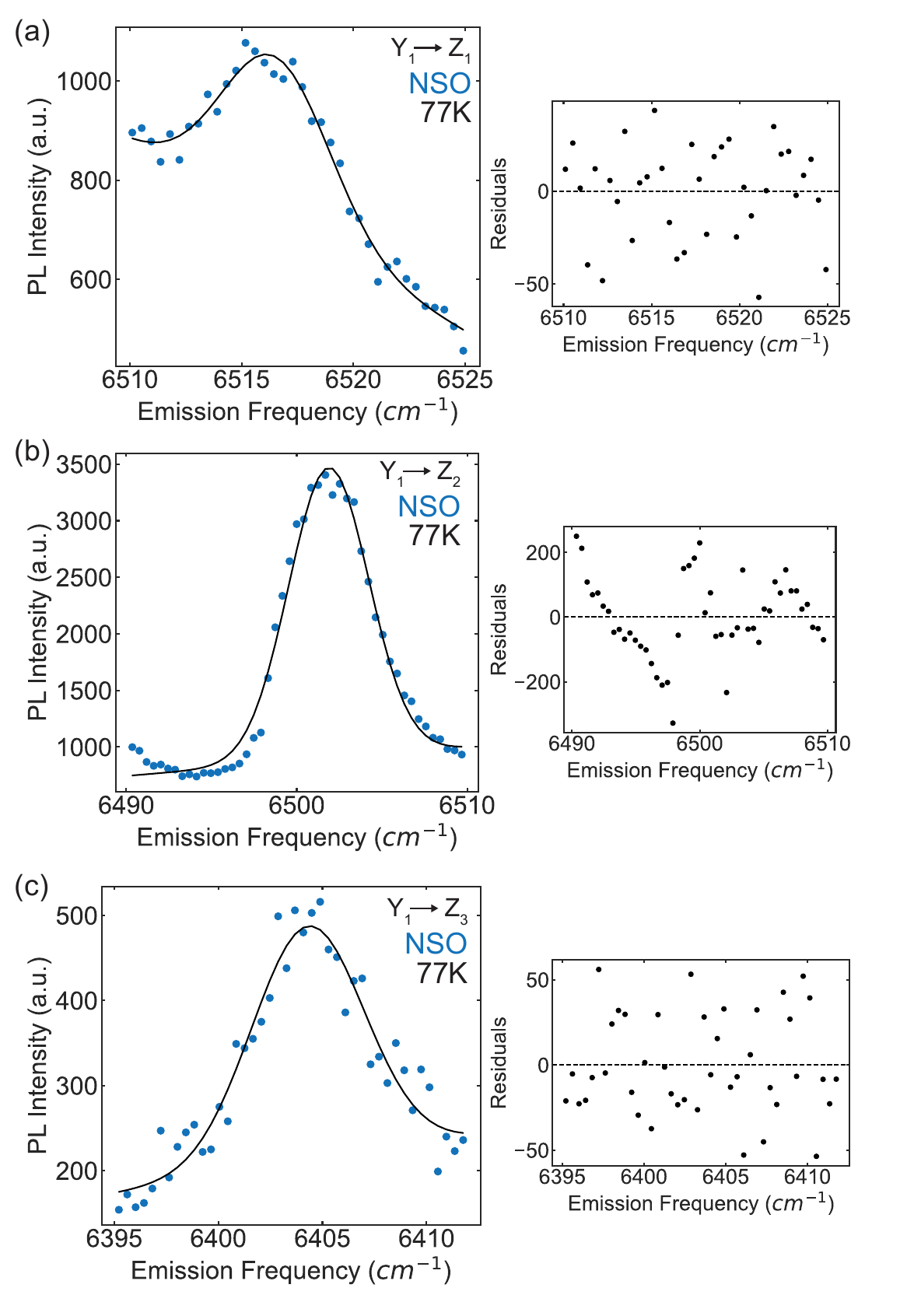}
    \caption{\textbf{Fits for \er-doped PTO on LSAT at 77\,K.} Gaussian fits for the \textbf{a} {$Y_{1} \rightarrow Z_{1}$}, \textbf{b} {$Y_{1} \rightarrow Z_{2}$}, and \textbf{c} {$Y_{1} \rightarrow Z_{3}$} transitions. Fits (black solid line) shown on top of data (dark blue circle). Corresponding residual from fits shown to the right of each plot. Sample was excited at 6500 \cm.}
    \label{NSO_77K}
\end{figure*}

Peak fits for the \er-doped PTO samples on LSAT (Fig. \ref{LSAT_77K}), STO (Fig. \ref{STO_77K}), DSO (Fig. \ref{DSO_77K}), GSO (Fig. \ref{GSO_77K}) and NSO (Fig. \ref{NSO_77K}) at 77\,K. Peaks were fit to a Gaussian curve with a linear background. The three peaks fitted correspond to the {$Y_{1} \rightarrow Z_{1}$} (Fig. \ref{LSAT_77K}a-\ref{NSO_77K}a), {$Y_{1} \rightarrow Z_{2}$} (Fig. \ref{LSAT_77K}b-\ref{NSO_77K}b), and  {$Y_{1} \rightarrow Z_{3}$} (Fig. \ref{LSAT_77K}c-\ref{NSO_77K}c) transitions. Residual of fits provided next to each peak fit to show the goodness of fits. Fit results are reported in Table \ref{tab:table3}.\\

\begin{table*}
\caption{Gaussian fit results of frequency (Freq.), intensity, and linewidth from Figures \ref{LSAT_77K}, \ref{STO_77K}, \ref{DSO_77K}, \ref{GSO_77K}, and \ref{NSO_77K}. Freq. from Literature (Lit.) from \cite{Stevenson2022Erbium-implantedApplications}.}
\label{tab:table3}
\begin{tabular}{cc|ccc|ccc}
\hline
 & & \multicolumn{3}{c}{LSAT}&\multicolumn{3}{c}{STO}\\
Transition & Freq. from Lit. & Freq. & Intensity & Linewidth & Freq. & Intensity & Linewidth\\
&(\cm) & (\cm) & (arb. units) & (\cm) & (\cm) & (arb. units) & (\cm)\\ \hline
{$Y_{1} \rightarrow Z_{1}$}& 6511.7 & 6515.82 & 2230.89 & 2.36 & 6515.81 & 3623.92 & 2.41\\
{$Y_{1} \rightarrow Z_{2}$}& 6497.49 & 6501.39 & 30101.09 & 2.2 & 6501.34 & 14655.47 & 2.19
\\
{$Y_{1} \rightarrow Z_{3}$}& 6398.23 & 6403.98 & 8918.9 & 2.42 & 6403.91 & 870.85 & 2.44 \\
\hline
\end{tabular}

\begin{flushleft}
\begin{tabular}{cc|ccc|ccc}
\hline
 & & \multicolumn{3}{c}{DSO}&\multicolumn{3}{c}{GSO}\\
Transition & Freq. from Lit. & Freq. & Intensity & Linewidth & Freq. & Intensity & Linewidth\\
&(\cm) & (\cm) & (arb. units) & (\cm) & (\cm) & (arb. units) & (\cm)\\ \hline
{$Y_{1} \rightarrow Z_{1}$} & 6511.7 & 6516.28 & 607.68 & 2.46 & 6516.5 & 238.42 & 2.57
\\
{$Y_{1} \rightarrow Z_{2}$}& 6497.49 & 6501.44 & 5405.47 & 2.13 & 6501.8 & 2839.2 & 2.25
\\
{$Y_{1} \rightarrow Z_{3}$}& 6398.23 & 6404.13 & 349.34 & 2.55 & 6404.47 & 254.66 & 2.58\\
\hline
\end{tabular}
\end{flushleft}

\begin{flushleft}
\begin{tabular}{cc|cccccc}
\hline
 & & \multicolumn{3}{c}{NSO}& & &\\
Transition & Freq. from Lit. & Freq. & Intensity & Linewidth & & &\\
&(\cm) & (\cm) & (arb. units) & (\cm) & & &\\ \hline
{$Y_{1} \rightarrow Z_{1}$} & 6511.7 & 6516.54 & 337.88 & 2.43 &    &    & \\
{$Y_{1} \rightarrow Z_{2}$} & 6497.49 &6501.87 & 2575.09 & 2.35 & & &\\
{$Y_{1} \rightarrow Z_{3}$} & 6398.23 & 6404.3 & 278.17 & 2.7 & & &\\
\hline
\end{tabular}
\end{flushleft}
\end{table*}

\section{\label{A:TDep}Temperature Dependent Photoluminescence}

\begin{figure*}
    \includegraphics[width=70mm]{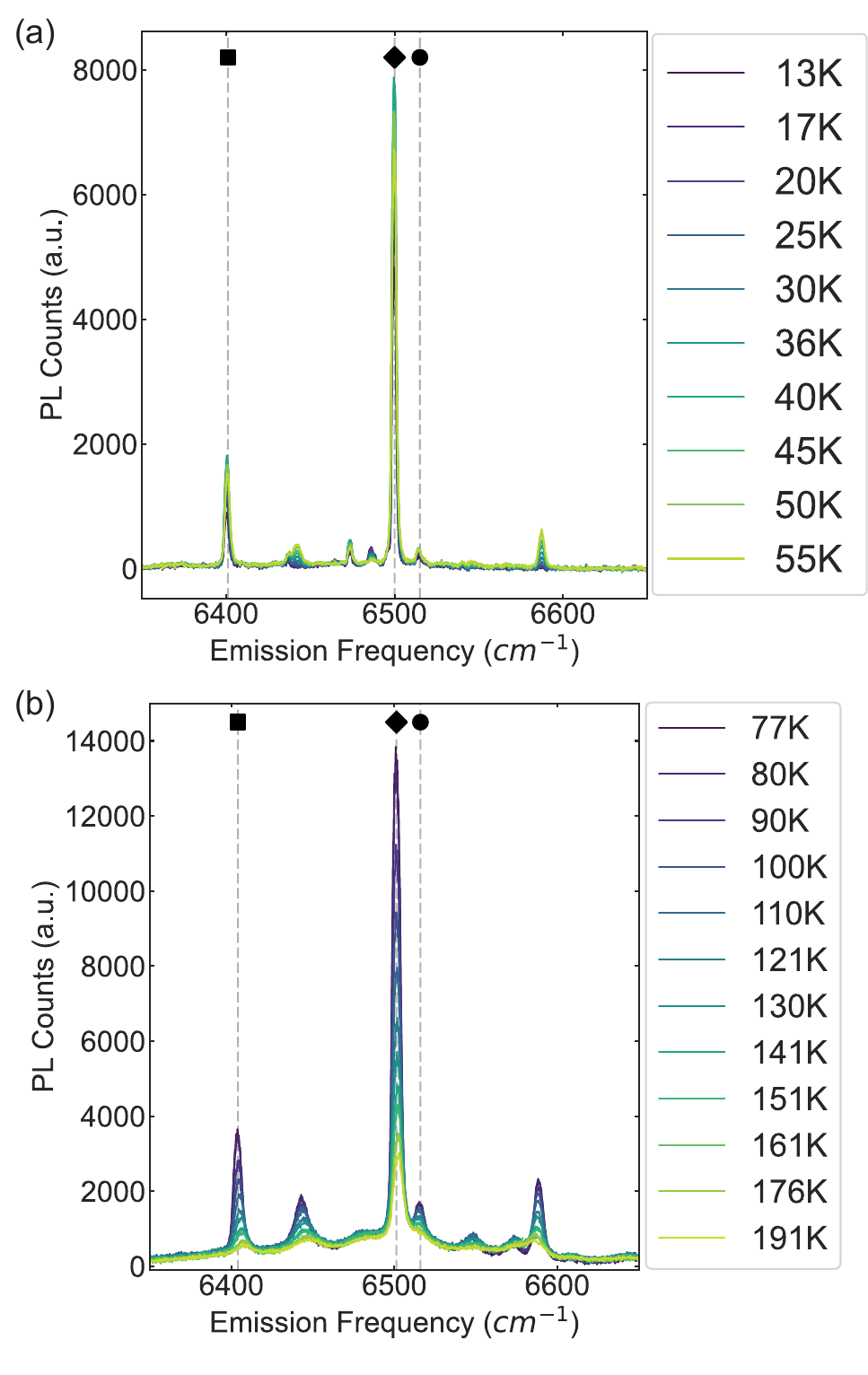}
    \caption{\textbf{Temperature Dependent Photoluminescence.} PL for \er-doped PTO on STO between \textbf{a} liquid He and liquid N$_2$ temperatures and between \textbf{b} liquid N$_2$ and room-temperature.}
    \label{Temp_PL}
\end{figure*}

\begin{figure*}
    \includegraphics[width=70mm]{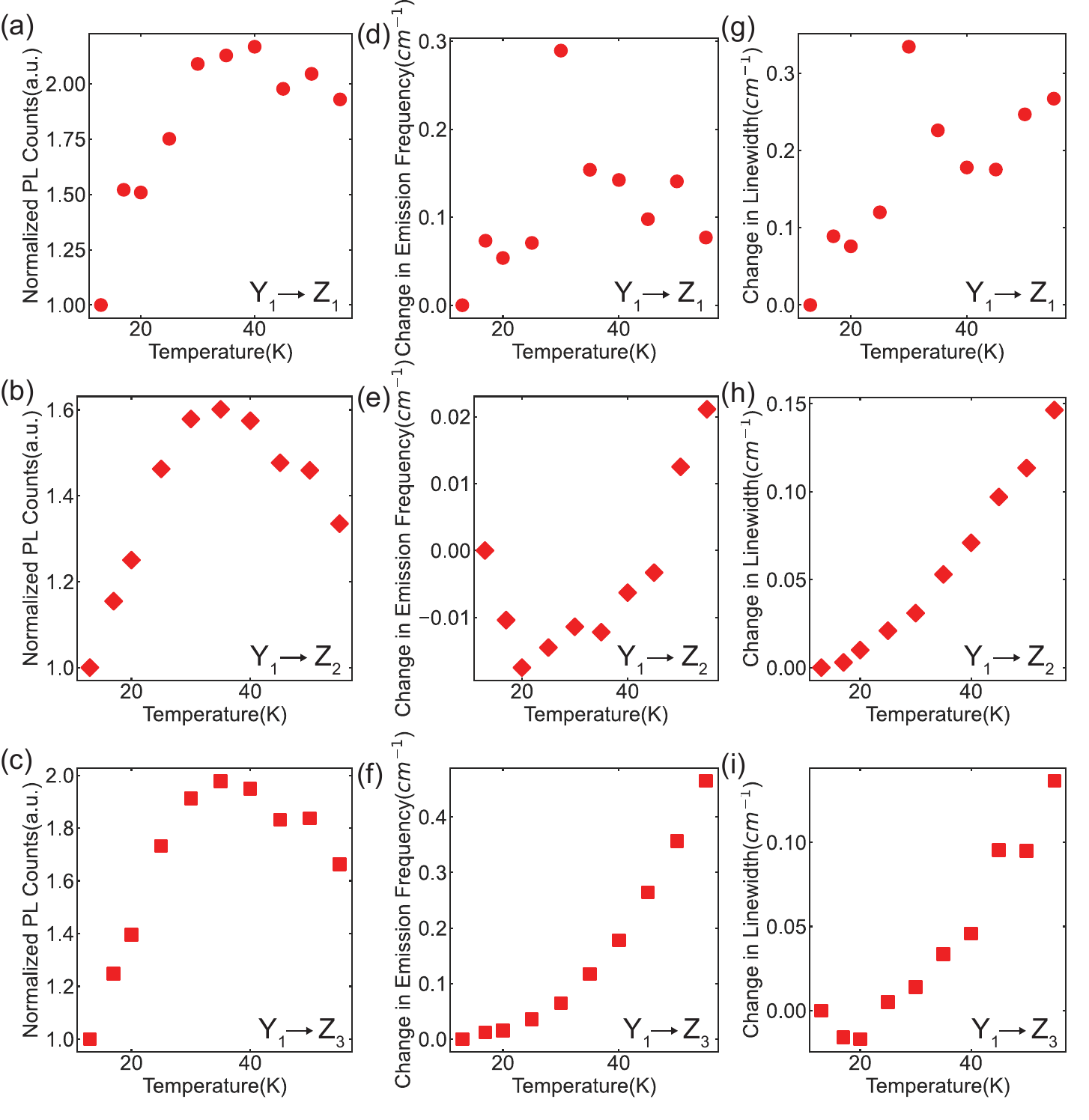}
    \caption{\textbf{Fit results from temperature dependent data at liquid He temperature regime.} Change in \textbf{a-c} PL counts, \textbf{d-f} emission frequency and \textbf{g-i} peak linewidth for the {$Y_{1} \rightarrow Z_{1}$}, {$Y_{1} \rightarrow Z_{2}$}, and {$Y_{1} \rightarrow Z_{3}$} transitions respectively at different temperatures. PL counts were normalized with respect to each peak at 13\,K. Change in emission frequency and linewidth were also determined with respect to each peak at 13\,K.}
    \label{Temp_LHe}
\end{figure*}

\begin{figure*}
    \includegraphics[width=70mm]{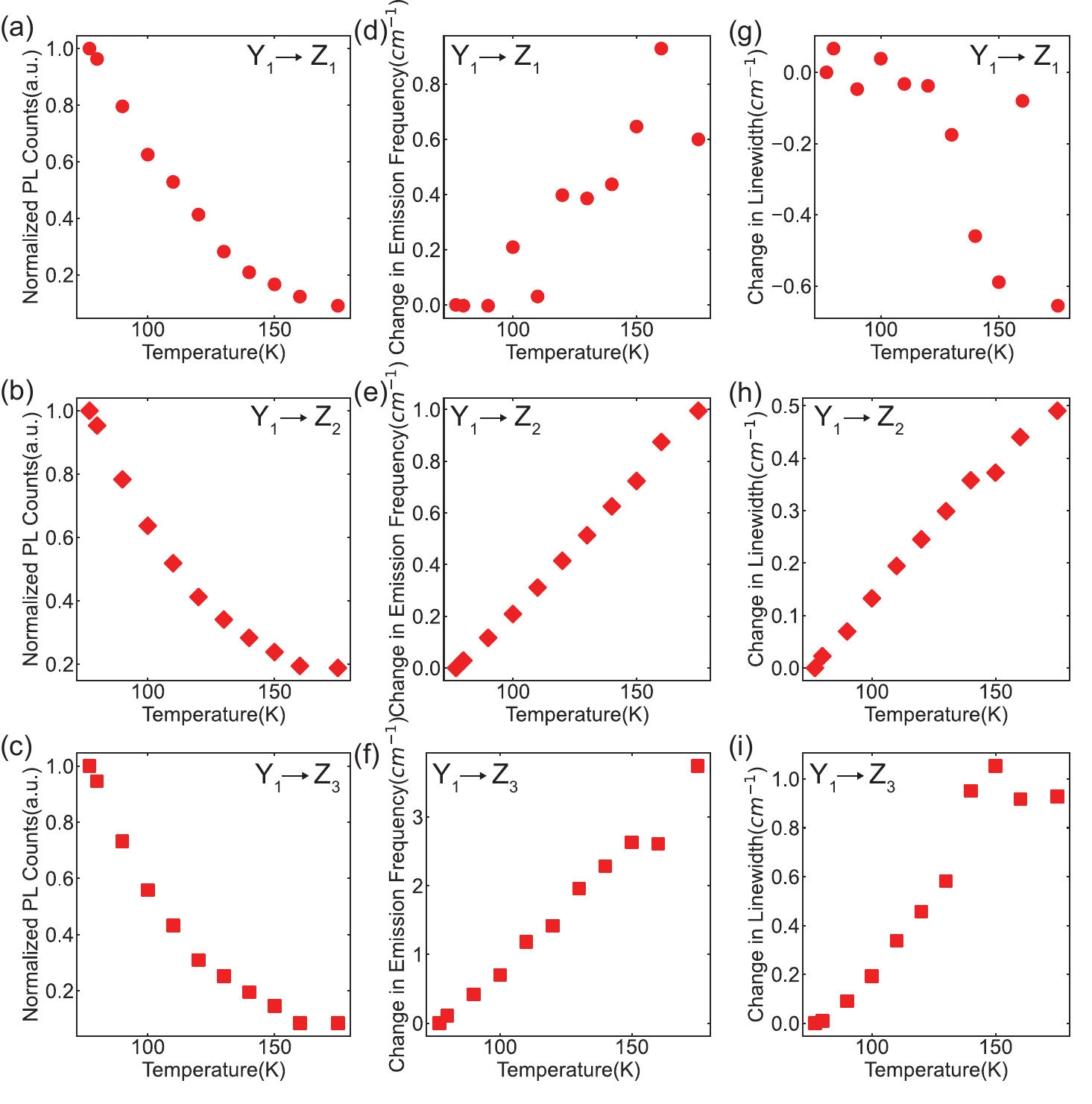}
    \caption{\textbf{Fit results from temperature dependent data at liquid N$_2$ temperature regime.} Change in \textbf{a-c} PL counts, \textbf{d-f} emission frequency and \textbf{g-i} peak linewidth for the {$Y_{1} \rightarrow Z_{1}$}, {$Y_{1} \rightarrow Z_{2}$}, and {$Y_{1} \rightarrow Z_{3}$} transitions respectively at different temperatures. PL counts were normalized with respect to each peak at 77\,K. Change in emission frequency and linewidth were also determined with respect to each peak at 77\,K.}
    \label{Temp_LN2}
\end{figure*}

Temperature dependent PL of \er-doped PTO on STO. Temperature dependent PL measurements were done for 2 temperature regimes: the liquid He temperature regime from 13\,K-55\,K (Fig. \ref{Temp_PL}a) and the liquid N$_2$ temperature regime from 77\,K-191\,K (Fig. \ref{Temp_PL}b). All spectra was measured with an excitation frequency of 6500 \cm. Peaks corresponding to {$Y_{1} \rightarrow Z_{1}$}, {$Y_{1} \rightarrow Z_{2}$}, and  {$Y_{1} \rightarrow Z_{3}$} transitions were fit to Gaussian peaks with a linear background. Peak fit results were plotted as a function of temperature for the three transitions in both the liquid He regime (Fig. \ref{Temp_LHe}) and the liquid N$_2$ regime (Fig. \ref{Temp_LN2}).\\

For the liquid He temperature regime, the normalized PL counts increase with temperature up until 35\,K and then begins to decrease for higher temperatures (Fig. \ref{Temp_LHe}a-c). At 35\,K, there is enough thermal energy to populate different states. Both the emission frequency (Fig. \ref{Temp_LHe}d-f) and linewidth (Fig. \ref{Temp_LHe}g-i) increase with temperature which is expected due to changes in the population of states with thermal energy. At most, a change in emission frequency of 0.47 \cm{} for the {$Y_{1}$} to {$Z_{3}$} transition and a change in linewidth of 0.27 \cm{} for the {$Y_{1}$} to {$Z_{1}$} transition are observed at 55\,K.\\

For the liquid N$_2$ temperature regime, the normalized PL counts (Fig. \ref{Temp_LN2}a-c) decreases with temperature while the change in emission frequency (Fig. \ref{Temp_LN2}d-f) and linewidth (Fig. \ref{Temp_LN2}g-i) increase. The only exception to this trend is the change in linewidth for the {$Y_{1} \rightarrow Z_{1}$} transition (Fig. \ref{Temp_LN2}g). The changes in emission frequency and linewidth observed in Fig. \ref{peak_comparison}g,h occur between 110\,K-150\,K and 90\,K-110\,K, respectively.  Therefore, the trends discussed in the \textit{Strain-Dependent Parameters} section are due to different structural distortions imposed by epitaxial strain on the thin film through the substrate and not because of differences in thermal conditions.\\

\section{Additional Peaks in \er-doped PTO on GSO and NSO}
\begin{figure*}
    \includegraphics[width=70mm]{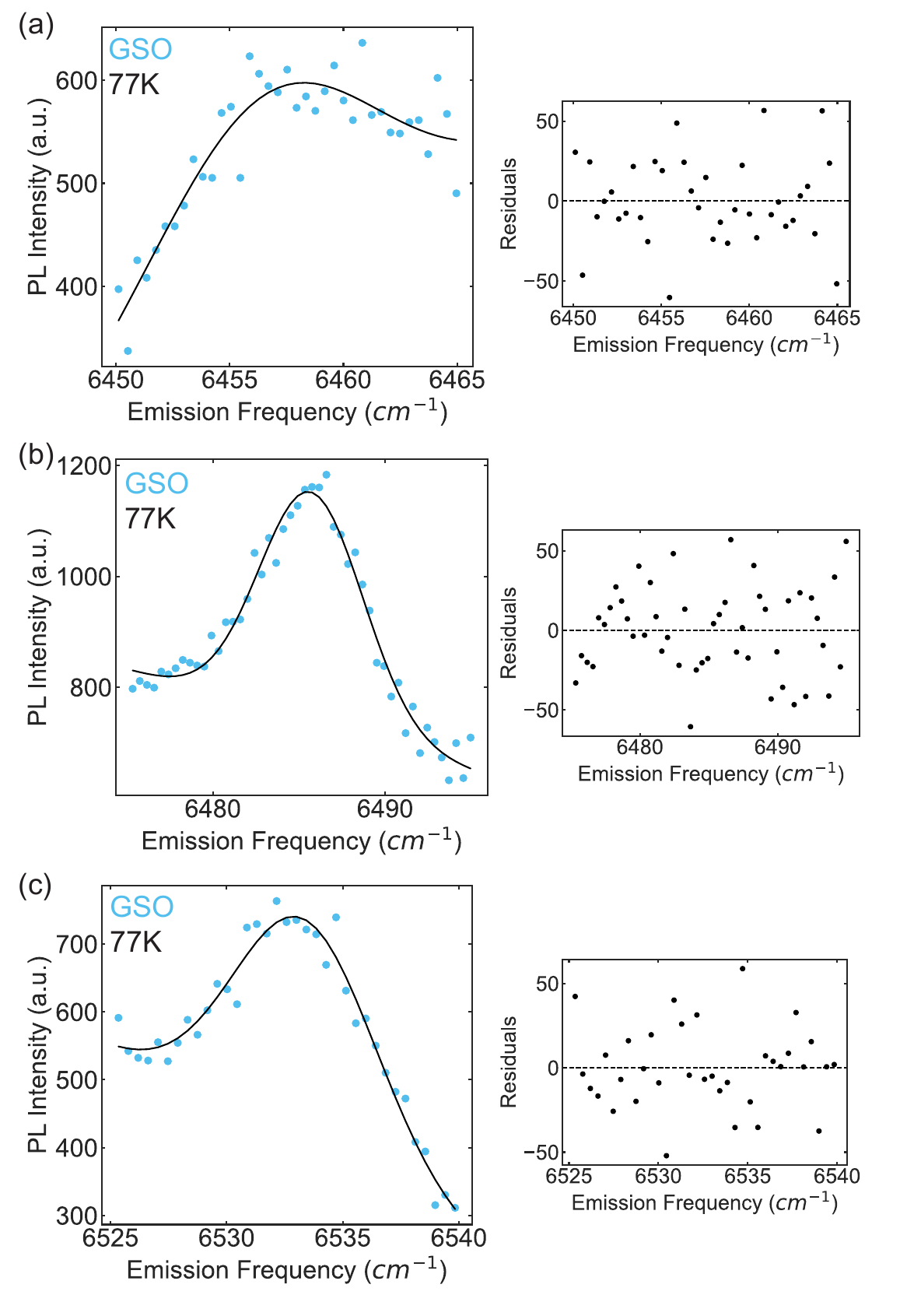}
    \caption{\textbf{Fits for additional peaks in \er-doped PTO on GSO at 77\,K.} Gaussian fits for additional peaks at \textbf{a} 6456 \cm, \textbf{b} 6485 \cm, and \textbf{c} 6533 \cm. Fits (black solid line) shown on top of data (light blue circle). Corresponding residual from fits shown to the right of each plot. Sample was excited at 6500 \cm.}
    \label{GSO_newpeaks}
\end{figure*}

\begin{figure*}
    \includegraphics[width=70mm]{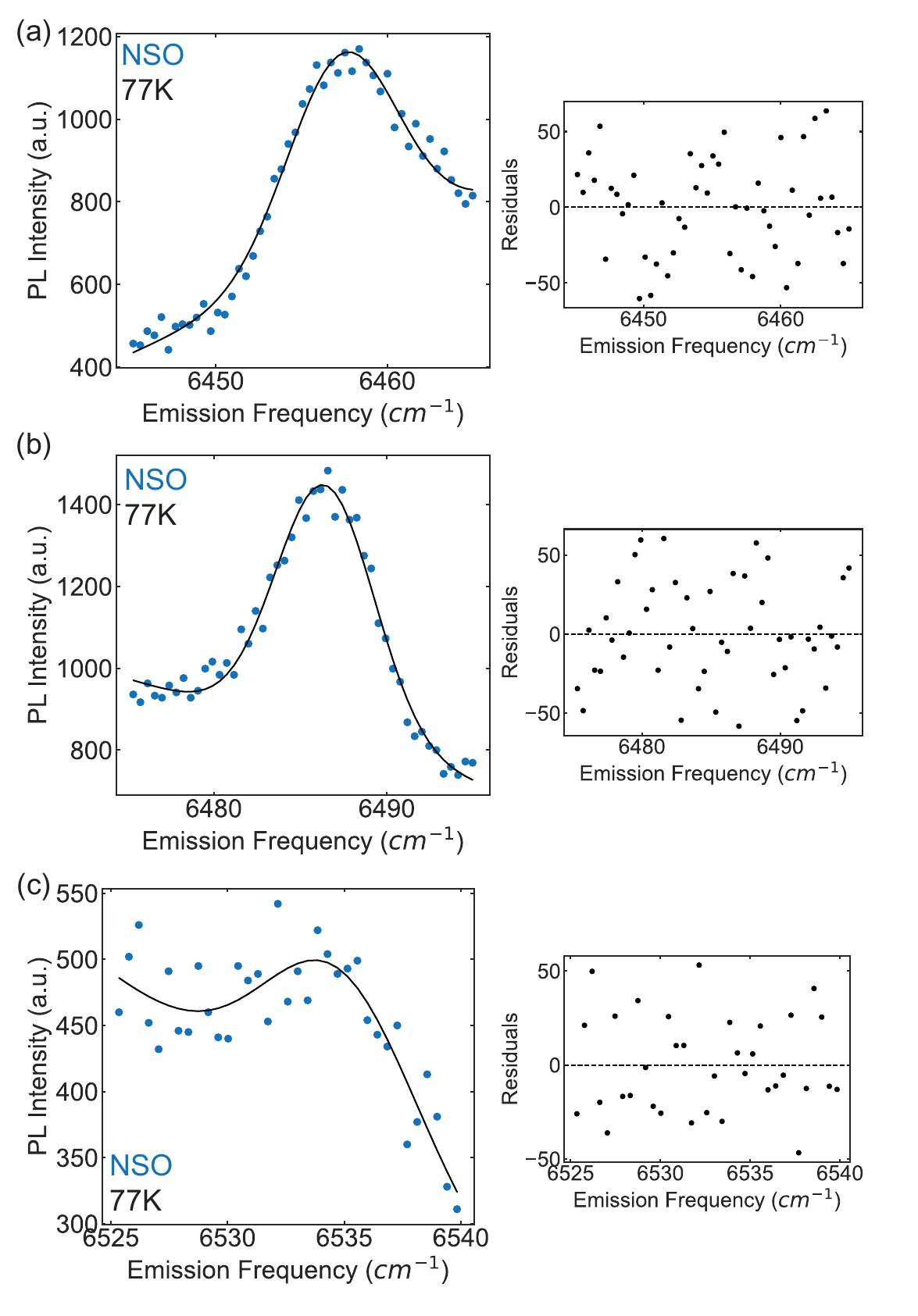}
    \caption{\textbf{Fits for additional peaks in \er-doped PTO on NSO at 77\,K.} Gaussian fits for additional peaks at \textbf{a} 6457 \cm, \textbf{b} 6486 \cm, and \textbf{c} 6535 \cm. Fits (black solid line) shown on top of data (dark blue circle). Corresponding residual from fits shown to the right of each plot. Sample was excited at 6500 \cm.}
    \label{NSO_newpeaks}
\end{figure*}

Peak fits for additional peaks in \er-doped PTO samples on GSO (Fig. \ref{GSO_newpeaks}) and NSO (Fig. \ref{NSO_newpeaks}) at 77\,K. Peaks were fit to a Gaussian curve with a linear background.  Residual of fits provided next to each peak fit to show the goodness of fits. Fit results are reported in Table \ref{tab:table1}.\\

\begin{figure*}
    \includegraphics[width=168mm]{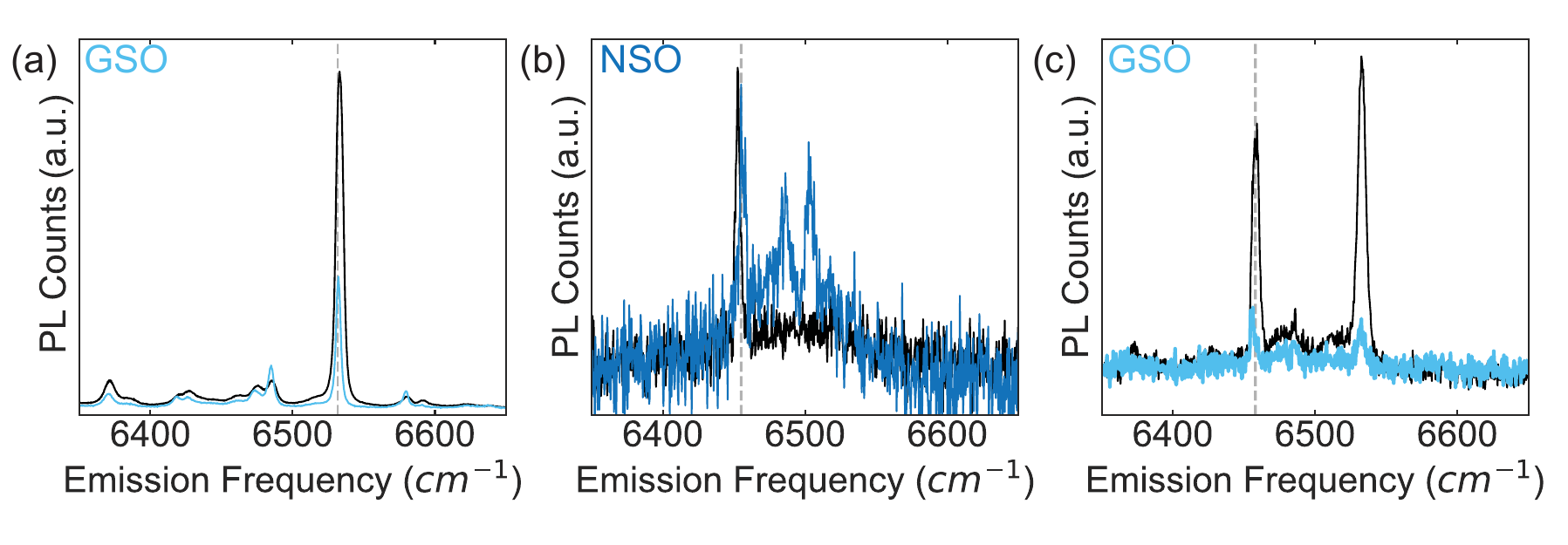}
    \caption{\textbf{Comparison of PL of \er-doped PTO and Substrates at Different Resonant Frequencies.} \textbf{a} PL of \er-doped PTO on GSO (light blue) and GSO substrate (black) excited at 6533 \cm, \textbf{b} PL of \er-doped PTO on NSO (dark blue) and NSO substrate (black) excited at 6456 \cm, and \textbf{c} PL of \er-doped PTO on GSO (light blue) and GSO substrate (black) excited at 6456 \cm. Dashed gray lines correspond to excitation frequency.}
    \label{REI_subs}
\end{figure*}

Additional comparisons of PL of \er-doped PTO samples and their respective substrates at resonant excitation frequencies to the additional peaks. In Figure \ref{REI_subs}a, \er-doped PTO on GSO (light blue) and GSO substrate (black) are excited at 6533 \cm{} (dashed line). At this excitation frequency, a different spectra is observed than at 6500 \cm{} excitation. The peaks at this excitation frequency are present for both sample and substrate. Since Gd$^{3+}$ does not have any transitions in this spectral range,\cite{Dieke1963TheEarths} we believe these peaks are due to \er{} impurities within the GSO substrate. On the other hand, in Figure \ref{REI_subs}b, when \er-doped PTO on NSO (dark blue) is excited at 6456 \cm{} (dashed line), peaks at 6456 \cm, 6486 \cm{} and 6500 \cm{} are observed. These peaks are also present at the 6500 \cm{} excitation. Additionally, no peaks are observed when the NSO substrate (black) is excited at 6456 \cm{} except for some counts at that frequency due to residual laser scattering. Lastly, in Figure \ref{REI_subs}c, \er-doped PTO on GSO (light blue) and GSO substrate (black) are excited at 6456 \cm{} (dashed line). Besides for the peak resonant to that excitation frequency, a peak at 6533 \cm{} is also observed for both the sample and substrate. The PL of the \er{} impurities in GSO are much brighter than the \er{} in PTO at this excitation frequency. Hence, we cannot differentiate if the source of those additional peaks are the \er{} dopants in PTO sample or the \er{} impurities in the GSO substrate.\\

\bibliography{main}% Produces the bibliography via BibTeX.

\end{document}